\documentclass[aps,prd,twocolumn,superscriptaddress,nofootinbib,showpacs]{revtex4-1}

\usepackage[pdftex]{graphicx}
\usepackage{epsfig}
\usepackage{amsmath}
\usepackage{amssymb}
\usepackage{dcolumn}% Align table columns on decimal point
\usepackage{multirow}

\newcommand\etal{{\it {et al., }}}

\begin{document}

\title{Multi--year search for dark matter annihilations in the Sun with the AMANDA--II and IceCube detectors}
\affiliation{III. Physikalisches Institut, RWTH Aachen University, D-52056 Aachen, Germany}
\affiliation{School of Chemistry \& Physics, University of Adelaide, Adelaide SA, 5005 Australia}
\affiliation{Dept.~of Physics and Astronomy, University of Alaska Anchorage, 3211 Providence Dr., Anchorage, AK 99508, USA}
\affiliation{CTSPS, Clark-Atlanta University, Atlanta, GA 30314, USA}
\affiliation{School of Physics and Center for Relativistic Astrophysics, Georgia Institute of Technology, Atlanta, GA 30332, USA}
\affiliation{Dept.~of Physics, Southern University, Baton Rouge, LA 70813, USA}
\affiliation{Dept.~of Physics, University of California, Berkeley, CA 94720, USA}
\affiliation{Lawrence Berkeley National Laboratory, Berkeley, CA 94720, USA}
\affiliation{Institut f\"ur Physik, Humboldt-Universit\"at zu Berlin, D-12489 Berlin, Germany}
\affiliation{Fakult\"at f\"ur Physik \& Astronomie, Ruhr-Universit\"at Bochum, D-44780 Bochum, Germany}
\affiliation{Physikalisches Institut, Universit\"at Bonn, Nussallee 12, D-53115 Bonn, Germany}
\affiliation{Dept.~of Physics, University of the West Indies, Cave Hill Campus, Bridgetown BB11000, Barbados}
\affiliation{Universit\'e Libre de Bruxelles, Science Faculty CP230, B-1050 Brussels, Belgium}
\affiliation{Vrije Universiteit Brussel, Dienst ELEM, B-1050 Brussels, Belgium}
\affiliation{Dept.~of Physics, Chiba University, Chiba 263-8522, Japan}
\affiliation{Dept.~of Physics and Astronomy, University of Canterbury, Private Bag 4800, Christchurch, New Zealand}
\affiliation{Dept.~of Physics, University of Maryland, College Park, MD 20742, USA}
\affiliation{Dept.~of Physics and Center for Cosmology and Astro-Particle Physics, Ohio State University, Columbus, OH 43210, USA}
\affiliation{Dept.~of Astronomy, Ohio State University, Columbus, OH 43210, USA}
\affiliation{Dept.~of Physics, TU Dortmund University, D-44221 Dortmund, Germany}
\affiliation{Dept.~of Physics, University of Alberta, Edmonton, Alberta, Canada T6G 2G7}
\affiliation{D\'epartement de physique nucl\'eaire et corpusculaire, Universit\'e de Gen\`eve, CH-1211 Gen\`eve, Switzerland}
\affiliation{Dept.~of Physics and Astronomy, University of Gent, B-9000 Gent, Belgium}
\affiliation{Max-Planck-Institut f\"ur Kernphysik, D-69177 Heidelberg, Germany}
\affiliation{Dept.~of Physics and Astronomy, University of California, Irvine, CA 92697, USA}
\affiliation{Laboratory for High Energy Physics, \'Ecole Polytechnique F\'ed\'erale, CH-1015 Lausanne, Switzerland}
\affiliation{Dept.~of Physics and Astronomy, University of Kansas, Lawrence, KS 66045, USA}
\affiliation{Dept.~of Astronomy, University of Wisconsin, Madison, WI 53706, USA}
\affiliation{Dept.~of Physics, University of Wisconsin, Madison, WI 53706, USA}
\affiliation{Institute of Physics, University of Mainz, Staudinger Weg 7, D-55099 Mainz, Germany}
\affiliation{Universit\'e de Mons, 7000 Mons, Belgium}
\affiliation{Bartol Research Institute and Department of Physics and Astronomy, University of Delaware, Newark, DE 19716, USA}
\affiliation{Dept.~of Physics, University of Oxford, 1 Keble Road, Oxford OX1 3NP, UK}
\affiliation{Dept.~of Physics, University of Wisconsin, River Falls, WI 54022, USA}
\affiliation{Oskar Klein Centre and Dept.~of Physics, Stockholm University, SE-10691 Stockholm, Sweden}
\affiliation{Department of Physics and Astronomy, Stony Brook University, Stony Brook, NY 11794-3800, USA}
\affiliation{Dept.~of Physics and Astronomy, University of Alabama, Tuscaloosa, AL 35487, USA}
\affiliation{Dept.~of Astronomy and Astrophysics, Pennsylvania State University, University Park, PA 16802, USA}
\affiliation{Dept.~of Physics, Pennsylvania State University, University Park, PA 16802, USA}
\affiliation{Dept.~of Physics and Astronomy, Uppsala University, Box 516, S-75120 Uppsala, Sweden}
\affiliation{Dept.~of Physics, University of Wuppertal, D-42119 Wuppertal, Germany}
\affiliation{DESY, D-15735 Zeuthen, Germany}

\author{R.~Abbasi}
\affiliation{Dept.~of Physics, University of Wisconsin, Madison, WI 53706, USA}
\author{Y.~Abdou}
\affiliation{Dept.~of Physics and Astronomy, University of Gent, B-9000 Gent, Belgium}
\author{T.~Abu-Zayyad}
\affiliation{Dept.~of Physics, University of Wisconsin, River Falls, WI 54022, USA}
\author{M.~Ackermann}
\affiliation{DESY, D-15735 Zeuthen, Germany}
\author{J.~Adams}
\affiliation{Dept.~of Physics and Astronomy, University of Canterbury, Private Bag 4800, Christchurch, New Zealand}
\author{J.~A.~Aguilar}
\affiliation{D\'epartement de physique nucl\'eaire et corpusculaire, Universit\'e de Gen\`eve, CH-1211 Gen\`eve, Switzerland}
\author{M.~Ahlers}
\affiliation{Dept.~of Physics, University of Wisconsin, Madison, WI 53706, USA}
\author{D.~Altmann}
\affiliation{III. Physikalisches Institut, RWTH Aachen University, D-52056 Aachen, Germany}
\author{K.~Andeen}
\affiliation{Dept.~of Physics, University of Wisconsin, Madison, WI 53706, USA}
\author{J.~Auffenberg}
\affiliation{Dept.~of Physics, University of Wisconsin, Madison, WI 53706, USA}
% ...

\collaboration{IceCube Collaboration}
\noaffiliation

\date{\today}
 
\begin{abstract}
  A search for an excess of muon--neutrinos from dark matter annihilations in the Sun has been performed 
with the AMANDA--II neutrino telescope using data collected in 812 days of livetime between 2001 and 2006 
and 149 days of livetime collected with the AMANDA--II and the 40--string configuration of IceCube during 
2008 and early 2009. No excess over the expected atmospheric neutrino background has been observed. 
We combine these results with the previously published IceCube limits obtained with data taken during 2007 
to obtain a total livetime of 1065 days. We provide an upper limit at 90\% confidence level on the 
annihilation rate of captured neutralinos in the Sun, as well as the corresponding muon flux limit at the 
Earth, both as functions of the neutralino mass in the range 50~GeV--5000~GeV. 
We also derive a limit on the neutralino--proton spin--dependent and spin--independent cross section. The 
limits presented here improve the previous results obtained by the collaboration 
between a factor of two and five, as well as extending the neutralino masses probed down to 50 GeV. 
The spin--dependent cross section limits are the most stringent so far for neutralino masses above 
200~GeV, and well below direct search results in the mass range from 50~GeV to 5~TeV.

\end{abstract}

%\keywords{}

\maketitle

\section{\label{sec:Intro} Introduction}

There is an impressive corpus of astrophysical and cosmological observations that indicate that a yet 
unknown form of non--luminous matter constitutes about 23\% of the content of the universe. 
This matter can not be baryonic though, as the abundance of baryons in any form is strongly constrained by 
the inferred primordially synthesized abundances of deuterium and helium, as well as the observed small-scale 
anisotropy in the cosmic microwave background (see, e.g., the review in Ref.~\cite{PDG:10}). However, the 
observational limits on non--baryonic dark matter in the form of relic stable particles from the Big Bang 
are not so strong and, indeed, a wealth of models exist that propose candidates with interaction cross 
sections, masses and relic densities which are compatible with observations. Generically, any particle 
physics model that provides a stable weakly--interacting massive particle (WIMP) is of interest from the 
point of view of the dark matter problem. One class of models that are extensively studied for their interest 
in particle physics are supersymmetric extensions of the Standard Model. Several flavours of supersymmetry 
are currently under scrutiny at the LHC, and both ATLAS and CMS have already probed parts of the parameter 
space of specific benchmark models~\cite{ATLAS,CMS}. 
In this paper we focus on the Minimal Supersymmetric extension to the Standard Model (MSSM),  
which provides a WIMP candidate in the lightest neutralino, $\widetilde{\chi}^{0}_{1}$, a 
linear combination of the supersymmetric partners of the electroweak neutral gauge-- and Higgs bosons. 
Assuming $R$--parity conservation, the neutralino is stable and a good dark matter candidate. Accelerator 
searches and relic density constraints from WMAP data allow a lower limit to be set  on the mass of 
the MSSM neutralino~\cite{Genieveve:04a}. Typical lower limits for $m_{\widetilde{\chi}^{0}_{1}}$ from such 
studies lie around  20 GeV, depending on the values chosen for tan$\beta$, the ratio of the  vacuum 
expectation values of the two neutral Higgses. Theoretical arguments based on the requirement of 
unitarity set an upper limit on  $m_{\widetilde{\chi}^{0}_{1}}$ of 340~TeV~\cite{Griest:90a}. Within these 
limits, the allowed parameter space of minimal supersymmetry can be exploited to build realistic models 
which provide relic neutralino densities of cosmological interest to address the dark matter problem. \par

Relic neutralinos in the galactic halo can lose energy through scatterings while traversing celestial bodies, like 
the Sun, and become gravitationally bound into orbits inside. The build-up of the accreted particles 
is limited by annihilations, which ultimately create high--energy neutrinos~\cite{Silk:85a}. 
In this paper we present two independent searches for a neutrino flux from the annihilations of 
neutralinos captured in the center of the Sun, performed with the AMANDA--II and IceCube neutrino 
telescopes. The overwhelming majority of triggers in AMANDA--II and IceCube, $\mathcal{O}$(10$^{10}$/year), 
are due to atmospheric muons reaching the depth of the array. The analyses are therefore based on the 
search for up--going muon tracks from the direction of the Sun when it is below the horizon, and have been 
optimized in order to maximize the sensitivity to the predicted signal from MSSM neutralinos. The results 
can, however, be interpreted  generically in terms of any other dark matter candidate which would produce 
a similar neutrino spectrum at the detector.
The first analysis (Analysis A) uses the data taken with AMANDA--II during 812 days of livetime in 
stand--alone operation between March 2001 and October 2006. In 2007 the detector was switched to a new 
data acquisition system (DAQ) which included full waveform recording, and this allowed us to integrate the 
detector as a subsystem of IceCube.  Analysis B uses 149 days of livetime collected between April 2008 
and April 2009 (when AMANDA--II was decommissioned) using the new AMANDA--II DAQ and data from both 
detectors, AMANDA--II and IceCube. \par

We have previously published a search for dark matter accumulated in the Sun using data taken in 2007 
with the 22--string configuration of IceCube~\cite{Abbasi:09b}. In Section~\ref{sec:combination} we 
combine these three independent analyses (Analysis A, Analysis B and the results from Ref.~\cite{Abbasi:09b}) which cover 
the period 2001--2008, presenting competitive limits on the muon flux from neutralino 
annihilations in the Sun and limits on the spin--dependent and spin--independent 
 neutralino--proton cross section.

\begin{figure}[t]
\includegraphics[width=\linewidth]{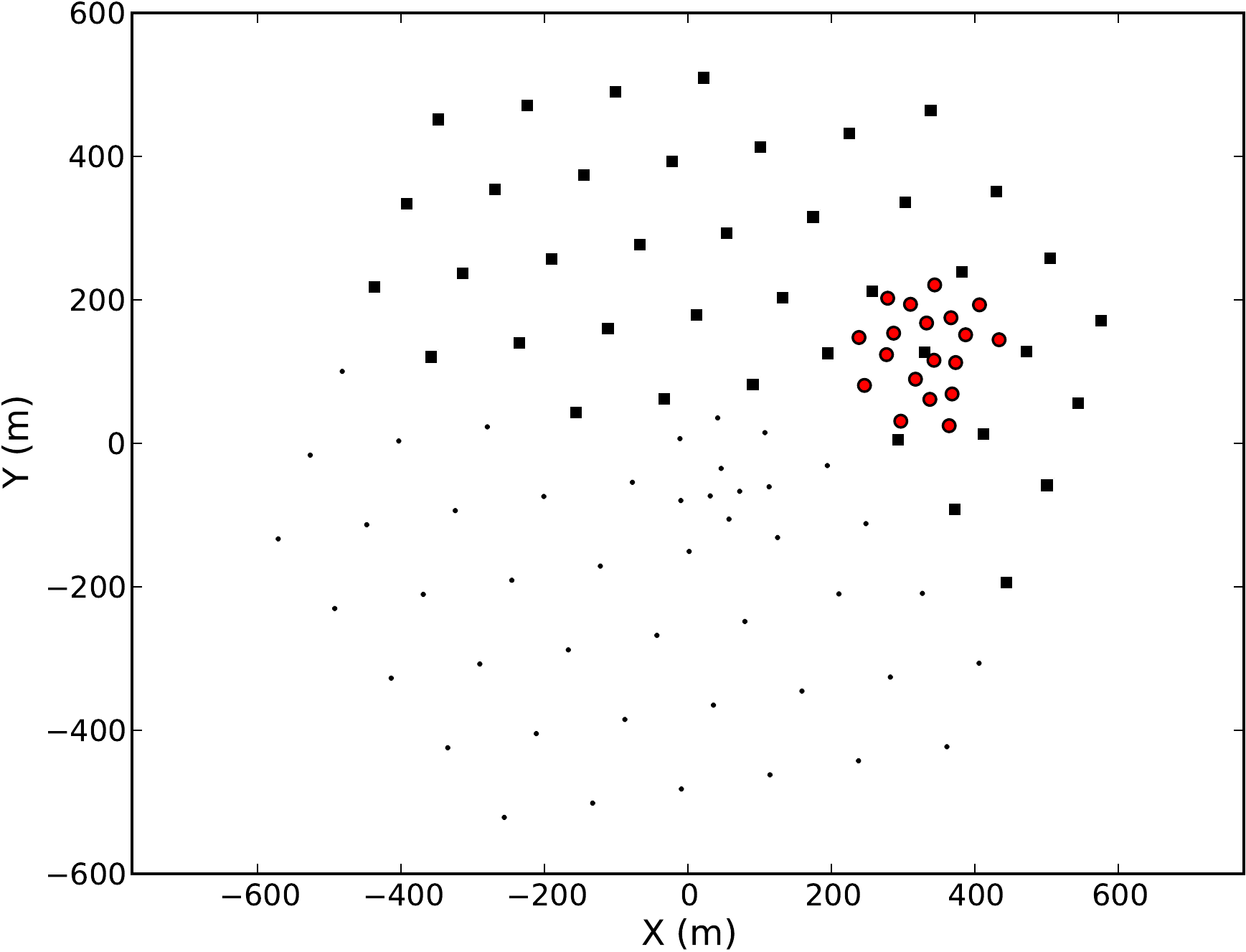}
\caption{Surface location of the IceCube 40--string configuration (filled squares) and AMANDA--II (filled circles)}
\label{fig:geometry}
\end{figure}

\section{\label{sec:detector}The AMANDA--II and IceCube detectors}
The AMANDA--II detector consisted of an array of 677 optical modules (OM) deployed on 19 vertical strings at depths 
between 1500~m and 2000~m in the South Pole ice cap. The optical modules consisted of a 20 cm diameter photomultiplier tube 
housed in a glass pressure vessel connected through a cable to the surface electronics, where the photomultiplier 
pulses were amplified, time--stamped and fed into the trigger logic. The inner ten strings used electrical analog signal 
transmission, while the outer nine strings  used optical fiber transmission. The strings were arranged in three approximately 
concentric circles of 60~m, 120~m and 200~m diameter respectively. The vertical separation of the optical modules in strings one to four 
was 20~m, and 10~m in strings five to nineteen. Several calibration devices (a 337 nm N$_2$ laser and three 
DC halogen lamps, one broadband and two with filters for 350 and 380 nm) were deployed at several locations in the array.  
Additionally, a YAG laser calibration system was set up on the surface, able to send light pulses through optical fibers to a diffuser 
located at each optical module. \par
The original AMANDA--II data acquisition system could only measure 
the maximum amplitude of the photomultiplier pulses and only eight hits could be buffered at a time. 
From 2003, each channel was connected to a Transient Waveform Recorder (TWR), a flash ADC that samples at 100 MHz with 12 bit 
resolution, being able to collect the full photomultiplier waveform in each channel~\cite{AMANDA:TWR}. The TWR system was run 
in parallel with the original DAQ until 2007, when the original DAQ was switched off. From this point, until its decommissioning in 
May 2009, AMANDA--II was run only with the TWR DAQ system.\par 
 The construction of the IceCube detector began in 2005 and 40 strings surrounding the 
AMANDA--II array had been deployed by 2008. With its denser string configuration, AMANDA--II played the role of a low--energy subdetector 
of IceCube, and both detectors where run in an 'OR' configuration until AMANDA--II was decommissioned. 
Figure~\ref{fig:geometry} shows the surface location of the AMANDA--II and IceCube strings during the 
2008 data taking period. The IceCube strings consist of 60 Digital Optical Modules (DOMs) separated vertically by 17~m. 
The inter--string separation is 125~m. Contrary to the AMANDA--II optical modules, the IceCube DOMs digitize the photomultiplier signals 
in--situ. They are also self--calibrating units, frequently exchanging timing calibration signals with the surface electronics. A description of the 
IceCube DOM and the DAQ can be found in Refs.~\cite{IceCube:PMT, IceCube:DAQ}. \par

Both IceCube and AMANDA--II used triggers that selected events based on the number of hit optical modules 
in a certain time window (at least 24 modules hit within 2.5 $\mu$s in AMANDA--II, and at least 
8 modules hit within 5 $\mu$s in IceCube). Additionally, a trigger tailored for low energy events required 
N hit modules in a given time window out of M consecutive modules in the same string (N/M condition).
 This trigger was set to 6/9 in strings one to four and 7/11 in strings five to nineteen for most 
of the AMANDA--II livetime, with a time window of 2.5 $\mu$s. The condition was 5/7 in all IceCube strings, 
with a time window of 1.5 $\mu$s.\par

  Muons from charged--current neutrino interactions near the array are detected by the Cherenkov light they produce when 
traversing the ice. The hit times along with the known detector geometry and the optical properties of the ice 
allow the reconstruction of the tracks passing through the detector. A  detailed description of the reconstruction techniques 
used in AMANDA and IceCube is given in Ref.~\cite{Ahrens:04a}\par

\section{\label{sec:sim} Signal and background simulations}
 The simulation of the neutralino--induced neutrino signal was performed using the {\texttt{WimpSim}} 
program~\cite{Blennow:08a} for a sample of neutralino masses (50, 100, 250, 500, 1000, 3000 and 5000 GeV).
Two extreme annihilation channels were considered in each case, a soft neutrino channel, 
$\widetilde{\chi}^{0}_{1}\widetilde{\chi}^{0}_{1}\rightarrow b\bar{b}$ and a hard neutrino channel, 
$\widetilde{\chi}^{0}_{1}\widetilde{\chi}^{0}_{1}\rightarrow W^{+}W^{-}$, 
($\widetilde{\chi}^{0}_{1}\widetilde{\chi}^{0}_{1}\rightarrow \tau^+\tau^-$ was chosen for 50 GeV 
neutralinos, since they are assumed to annihilate at rest and this mass is therefore below the $W$ 
production threshold).  This choice covers the range of neutrino energies that would be detectable 
with AMANDA--II/IceCube for typical MSSM models. Note, however, that neutrinos with energies above about 
a few hundred GeV will interact in the Sun, and do not escape; only the lower energy neutrinos from 
the decays of the products of such interactions will get out of the dense solar interior. 
This cutoff, rather than the WIMP mass, sets the neutrino spectrum for WIMP masses above 1 TeV. 
The simulated angular range was restricted to zenith angles between 90$^\circ$ (horizontal)
and 113$^\circ$, with the generated number of events as a function of angle weighted by the time the Sun spends at each 
declination. {\texttt{WimpSim}} propagates the neutrinos  taking into account energy losses and oscillations in their way 
out of the Sun as well as vacuum oscillations to the Earth, giving the expected neutrino flux at the location of the detector 
for each neutralino mass simulated. The neutrino--nucleon interactions in the ice around the detector producing the detectable 
muon flux were simulated with {\texttt{nusigma}}~\cite{Blennow:08a} using the {\texttt{CTEQ6}}~\cite{CTEQ6} parametrization of 
the nucleon structure functions.\par

The background for the dark matter searches arises from up--going atmospheric neutrinos and mis--reconstructed downward--going 
atmospheric muons. We have simulated the atmospheric neutrino flux according to Ref.~\cite{Lipari:93a}, using the {\texttt{ANIS}} program~\cite{ANIS}, 
with energies between 10~GeV and 325~TeV and zenith angles between 80$^\circ$ and 180$^\circ$ (vertically up--going).
The simulation includes neutrino propagation through the Earth, taking into account the Earth density profile~\cite{Dziewonski:81a}, neutrino absorption 
and neutral current scattering. The simulation of atmospheric muons was based on the {\texttt{CORSIKA}} air shower generator~\cite{CORSIKA} 
using the South Pole atmosphere parameters and the H\"orandel parametrization of the cosmic ray composition~\cite{Horandel:03a}.
 We have simulated 10$^{11}$ interactions, distributed uniformly with zenith angles between 0$^\circ$ and 90$^\circ$,
and with primary energies, $E_p$, between 600~GeV and 10$^{11}$~GeV. We note that the background simulations were not used in the 
evaluation of the actual background remaining at final cut level in the analyses presented below, but off--source data was 
used to that end. Background simulations were used as a consistency check at the different steps of the analyses.\par

Muons were propagated from the production point to the  detector taking into account energy losses by bremsstrahlung, 
pair production, photo--nuclear interactions and $\delta$--ray production as implemented in the code {\texttt{MMC}}~\cite{MMC}. 
The Cherenkov light produced by the muon tracks and secondaries was propagated to the optical modules taking into account photon scattering 
and absorption according to the measured optical properties of the ice at the detector~\cite{Askebjer:95a}, as well as the measured dust layer 
structure of the ice at the South Pole~\cite{Bramall:05a}. The {\texttt{Photonics}}~\cite{Photonics} program was used to this end.

\begin{table*}[t]
\begin{ruledtabular}
\begin{tabular}{ccccccccc}  
  $m_{\widetilde{\chi}^{0}_{1}}$(GeV) & channel & 
  ${\bar{\Psi}}$(deg)& $V_{\mathrm{eff}}$ (m$^3$) & $\mu_{s}^{90}$ & $\Gamma_{\mathrm A}$(s$^{-1}$)
    & $\Phi_{\mu}$ (km$^{-2}$ y$^{-1}$)   &  $\sigma_{\chi p}^{\mathrm{SD}}$  (cm$^2$) & $\sigma_{\chi p}^{\mathrm{SI}}$ (cm$^2$) \\
    &  & &   & &   & &  & \\\hline                                                          
  50    & $\tau^+\tau^-$ &  6.8 & 2.4$\times$10$^{4}$& 16.4 &  1.2$\times$10$^{24}$ & 2.8$\times$10$^4$ & 2.66$\times$10$^{-39}$ & 1.02$\times$10$^{-41}$\\
         & $b\bar{b}$   &  13.1 &  3.6$\times$10$^{3}$& 31.7 & 4.5$\times$10$^{26}$ & 4.7$\times$10$^5$ & 7.70$\times$10$^{-37}$ &2.94$\times$10$^{-39}$  \\\hline
  100    & $W^+W^-$     &  4.6 &  2.7$\times$10$^{5}$& 16.5 &  1.1$\times$10$^{23}$ & 3.9$\times$10$^3$ & 8.67$\times$10$^{-40}$ & 1.94$\times$10$^{-42}$\\
         & $b\bar{b}$   &  6.8 &  3.0$\times$10$^{4}$& 21.9 & 1.0$\times$10$^{25}$ & 3.2$\times$10$^4$ & 7.27$\times$10$^{-38}$ & 1.62$\times$10$^{-40}$\\\hline
  250    & $W^+W^-$     &  3.5 &  2.1$\times$10$^{6}$& 15.2 & 6.9$\times$10$^{21}$ & 1.0$\times$10$^3$ & 3.24$\times$10$^{-40}$ & 4.05$\times$10$^{-43}$\\
         & $b\bar{b}$   &  4.4 &  1.2$\times$10$^{5}$& [2.3,22.5] & 5.6$\times$10$^{23}$ & 6.5$\times$10$^3$ & 2.12$\times$10$^{-38}$ & 2.65$\times$10$^{-41}$\\ \hline
  500    & $W^+W^-$     &  3.2 &  4.1$\times$10$^{6}$& 13.5 & 2.6$\times$10$^{21}$ & 6.9$\times$10$^2$ & 4.85$\times$10$^{-40}$ & 4.47$\times$10$^{-43}$\\
         & $b\bar{b}$   &  3.9 &  6.2$\times$10$^{5}$& 11.4 & 5.5$\times$10$^{22}$ & 1.4$\times$10$^3$ & 9.81$\times$10$^{-39}$ & 9.05$\times$10$^{-42}$\\ \hline
  1000    & $W^+W^-$    &  3.0 &  5.3$\times$10$^{6}$& 7.4 & 1.2$\times$10$^{21}$ & 3.6$\times$10$^3$ & 8.82$\times$10$^{-40}$ & 6.81$\times$10$^{-43}$\\
         & $b\bar{b}$   &  3.6 &  1.1$\times$10$^{5}$& 14.2 &  2.7$\times$10$^{22}$ & 1.2$\times$10$^3$ & 1.86$\times$10$^{-38}$ & 1.44$\times$10$^{-41}$\\ \hline
  3000    & $W^+W^-$    &  3.0 &  5.7$\times$10$^{6}$& 11.2 & 2.3$\times$10$^{21}$ & 5.1$\times$10$^2$ & 1.53$\times$10$^{-38}$ &  1.05$\times$10$^{-41}$\\
         & $b\bar{b}$  &  3.4 &  1.7$\times$10$^{6}$& 10.7 & 9.6$\times$10$^{21}$ & 6.9$\times$10$^2$ & 6.16$\times$10$^{-38}$ &4.21$\times$10$^{-41}$ \\ \hline
  5000    & $W^+W^-$   &  3.0 &  5.4$\times$10$^{6}$& 9.5& 1.9$\times$10$^{22}$ & 4.3$\times$10$^2$ & 4.38$\times$10$^{-38}$ & 2.92$\times$10$^{-41}$\\
         & $b\bar{b}$  &  3.4 &  1.9$\times$10$^{6}$& 10.3 & 8.0$\times$10$^{21}$ & 6.5$\times$10$^2$ & 1.43$\times$10$^{-37}$ & 9.55$\times$10$^{-41}$ \\ 
\end{tabular}
\end{ruledtabular}
\caption{\label{tab:results_A} ANALYSIS A: For each neutralino mass and annihilation channel the 
table shows: The median of the space--angle distribution, ${\bar{\Psi}}$,  the effective volume, 
V$_{\mathrm{eff}}$,  the 90\% CL upper limit on the expected signal, $\mu_{\mathrm s}^{90}$, and the corresponding 90\% CL limits on the annihilation rate at the center of the Sun, 
$\Gamma_{\mathrm{A}}$, on the muon flux at the Earth, $\Phi_{\mu}$, and on the spin--dependent and spin--independent neutralino--proton cross sections, 
$\sigma_{\chi p}^{\mathrm{SD}}$  and  $\sigma_{\chi p}^{\mathrm{SI}}$. The limits include systematic uncertainties.}
\end{table*}

\section{\label{sec:analysis} Data analysis}

 Below we describe the two analyses performed with AMANDA--II data in stand--alone mode and with AMANDA--II run in coincidence with the 40--string 
configuration of IceCube. In both cases the cuts were optimized on data when the Sun was above the horizon, so the data analysis was kept blind to 
the actual direction of the Sun. 

\subsection{\label{subsec:analysis_A} ANALYSIS A\\ (AMANDA--only analysis)}
 The data set used in this analysis corresponds to a total of 812 days of effective livetime, and comprises a total of 7.25$\times$10$^9$ events collected when 
the Sun is below the horizon between the beginning of March 2001 and the end of October 2006. 
Data from periods where the detector showed unstable behaviour or periods where test or calibration runs were performed, were removed from the 
final data sample. 
The remaining events were cleaned of hits induced by electronic cross--talk, dark noise or unstable modules, and the data set was retriggered  
to make sure that the required amount of physical hits participated in building the trigger. The events were first reconstructed 
with two fast first--guess track finding algorithms, DirectWalk~\cite{Ahrens:04a} and JAMS~\cite{Ackermann:06c}. These reconstructions are aimed 
at identifying muon tracks and give a rough first estimate of their direction, using it as a first angular cut in order 
to reduce the data sample, still dominated by down--going atmospheric muons at this level. Events with zenith angles smaller than 70$^\circ$ as reconstructed by the 
DirectWalk or the JAMS reconstructions were rejected at this stage.  
 Two maximum likelihood reconstructions were applied to the remaining events. One uses 
DirectWalk and JAMS as seeds, performing an iterative maximization of the probability of observing the actual event geometry 
(hit times and positions) with respect to a given track direction. In reality, the negative of the logarithm of the likelihood is 
minimized in order to find the best--fit zenith and azimuth. In order to exploit the fact that most of the events that trigger the 
detector are down--going atmospheric muons, an iterative Bayesian reconstruction incorporating the known atmospheric muon zenith angle 
distribution as prior, was also performed. A comparison between the likelihoods of the standard and the Bayesian reconstructions can then be 
used to evaluate the likelihood that an event is down--going or up--going.\par
 The reconstructed events were then processed through a series of more stringent angular cuts on the direction of the tracks as obtained 
with the likelihood fits (zenith angle $>$ 80$^\circ$), to reduce the atmospheric muon background and retain as much of a potential signal as possible. 
At this level, only a fraction of 3$\times$10$^{-3}$ of the data and the simulated atmospheric muon background survive, while 69\% of the simulated 
atmospheric neutrino background was kept. Between 36\% and 78\% of the neutralino signal survived these cuts, depending on the neutralino mass and 
annihilation channel.\par

 The final event selection step was performed using a boosted decision tree (BDT) classifier~\cite{BDT}. 
%starting from the data sample to be separated, a sequence of binary splits is performed, using the variable that best separates between signal and background 
%at each given step. The same variable may be used at several steps, while other variables might be used more seldom. This allows the variables to be ranked 
%according to their usefulness in the classification task, and the tree to be pruned of variables with low separation capability. 
We have  selected 21 variables that showed good separation power and a correlation below 65\% between any pair. The variables used for the classification 
scheme can be grouped in two major classes: variables related to the hit topology of the event and variables related to the quality of the track reconstruction. 
Among the first class there are variables like the number of hit optical modules, the number of non--scattered, or 'direct', hits\footnote{Hits with a time stamp that 
corresponds to the time that light takes to travel directly from the track hypothesis to the optical module.}, the number of strings with hits within a 50 m radius cylinder 
around the track, and the number of strings with direct hits in the same cylinder, the center of gravity (c.o.g.) of the spatial position hits, the distance of 
the c.o.g. to the geometrical center of the detector, the length of the direct hits projected onto the direction of the track, the smoothness of the distribution 
of direct hits along the track and variables related to the probability of detecting a photon in the hit modules given the reconstructed track hypothesis. 
Among the variables related to the reconstruction quality, we have used the angle of the standard maximum likelihood reconstruction, the difference of 
log--likelihoods between the standard reconstruction and the Bayesian reconstruction and a measure of the angular resolution of the first--guess reconstructions. \par

In order to exploit the differences in the final muon energy spectra at the detector produced by the annihilation of neutralinos of different mass, 
we trained BDTs separately for each neutralino mass and annihilation channel, using the same 21 variables in each case. The 
signal training samples consisted of  50\% of each of the simulated signal samples. Data when the Sun is above the horizon were 
used as the background sample. The BDT classifies the given data as background or signal according to a continuous parameter that takes values between 
1 (signal--like) and --1 (background--like). A cut on the BDT output was chosen for each model as to maximize the discovery potential. 
Details of the data analysis and a complete list of the variables used can be found in Ref.~\cite{Alfio:10a}.\par 
 
 After the BDT cut the data sample is reduced by a factor between 1$\times$10$^{-7}$ and 3$\times$10$^{-7}$ with respect to trigger level, depending on 
the signal model used for the optimization, and its purity is closer to the irreducible atmospheric neutrino background. We note however that 
the approach taken in this and the analysis described in Section~\ref{subsec:analysis_b} does not require to reach a pure atmospheric neutrino sample,  
since we use the shape of the normalized space angular distributions with respect to the Sun of both signal and background to build our hypothesis 
testing. Details are given in Section~\ref{sec:results} below.\par

\subsubsection{\label{subsec:sys_A}Systematic uncertainties}
  The effect of different systematic uncertainties on the signal expectation was evaluated by varying the relevant parameters 
in the Monte Carlo and processing this new sample through the same analysis chain as the nominal sample. Systematic uncertainties affect the sensitivity 
of the detector to a given signal, a quantity that is characterized by the effective volume. The effective volume is defined as $V_{\mathrm{eff}}\,=\,(n_{\mathrm{final}}/n_{\mathrm{gen}})\times V_{\mathrm{gen}}$ 
and calculated by generating a given number of signal events $n_{\mathrm{gen}}$ in a geometrical volume $V_{\mathrm{gen}}$ around the detector. $n_{\mathrm{final}}$ is the remaining number of 
events after the analysis cuts have been applied to the generated event sample. The effective volume depends on neutrino energy, but in what follows we will 
show the integrated volume over the neutrino spectrum produced by a given neutralino mass. The relative uncertainty in the effective volume is then given by, 
\begin{equation}
\frac{\Delta V_{\mathrm{eff}}}{V_{\mathrm{eff}}} = \frac{V_{\mathrm{sys}}-V_{\mathrm{eff}}} {V_{\mathrm{eff}}}
\label{eq:veff}
\end{equation}
where $V_{\mathrm{eff}}$ is the effective volume of the baseline analysis and $V_{\mathrm{sys}}$ the effective volume calculated with a 
given assumption for systematic effects. Since we are not relying on atmospheric muon and neutrino Monte Carlo to estimate the background, we 
evaluate the effect of systematics only on the signal expectation.\par

 Systematic uncertainties can be classified in different categories. There are systematics induced from the uncertainties in quantities used in the 
signal Monte Carlo; neutrino cross sections, oscillation parameters or muon energy losses. The uncertainties in the oscillation parameters used to calculate 
the expected neutrino flux at the detector were taken from Ref.~\cite{Maltoni:04a} and lead to an uncertainty of less than $\pm$3\%. 
Further, uncertainties from the neutrino--nucleon cross section calculation within {\texttt{WimpSim}} and the simulation of muon energy loss in the ice have 
been estimated to be $\pm$7\%~\cite{Blennow:08a} and $\pm$1\%~\cite{MMC} respectively.\par
 An additional source of systematics is the implementation of the optical properties of the ice in the detector response 
simulation, as well as the in--situ optical module sensitivity. The AMANDA--II calibration light sources were used to measure photon arrival time distributions 
as a function of the relative emitter--receiver distance. Such measurements allow us, in principle, to extract an effective scattering length and absorption 
length that characterize the deep ice where the detector is located. However, the ice at the South Pole presents a layered structure, 
with slightly different optical properties due to the presence of different concentrations of dust at different depths~\cite{Ackermann:06b}.  
Moreover the process of melting the ice to deploy the strings and the subsequent refreeze of the water column changes the local optical properties 
of the ice in the drill hole. The inverse problem of extracting the ice properties from the measured photon arrival time information is then a quite 
difficult one, and different improved implementations have been applied in AMANDA with time. We have used the two most recent ice models as an estimation 
of the uncertainties introduced by the different implementations of the ice properties as a function of depth. The relative uncertainty induced in the detector 
effective volume by this effect ranges between 3\% and 30\%, depending on neutrino energy. \par
 The uncertainty on the total sensitivity of the deployed optical modules (glass plus PMT) also contributes to the uncertainty 
in the effective volume. Two additional Monte Carlo samples with the light collection efficiency of each module globally shifted 
by $\pm$10\% with respect to the baseline simulation were produced and the events passed through the complete the analysis chain. The effect 
on the effective volume lies between 20\% and 40\%.  The overall systematic uncertainties in the detector 
effective volume lie between 20\% and about 50\% depending on the neutralino mass model being tested. The total uncertainty in 
$V_{\rm eff}$ has been calculated under the assumption that all uncertainties are uncorrelated, i.e., $\Delta V/V\,=\,\sqrt{\sum_{i} (\Delta V/V)_{i}^2}$

\begin{figure}[t]
\includegraphics[height=0.75\linewidth,width=\linewidth]{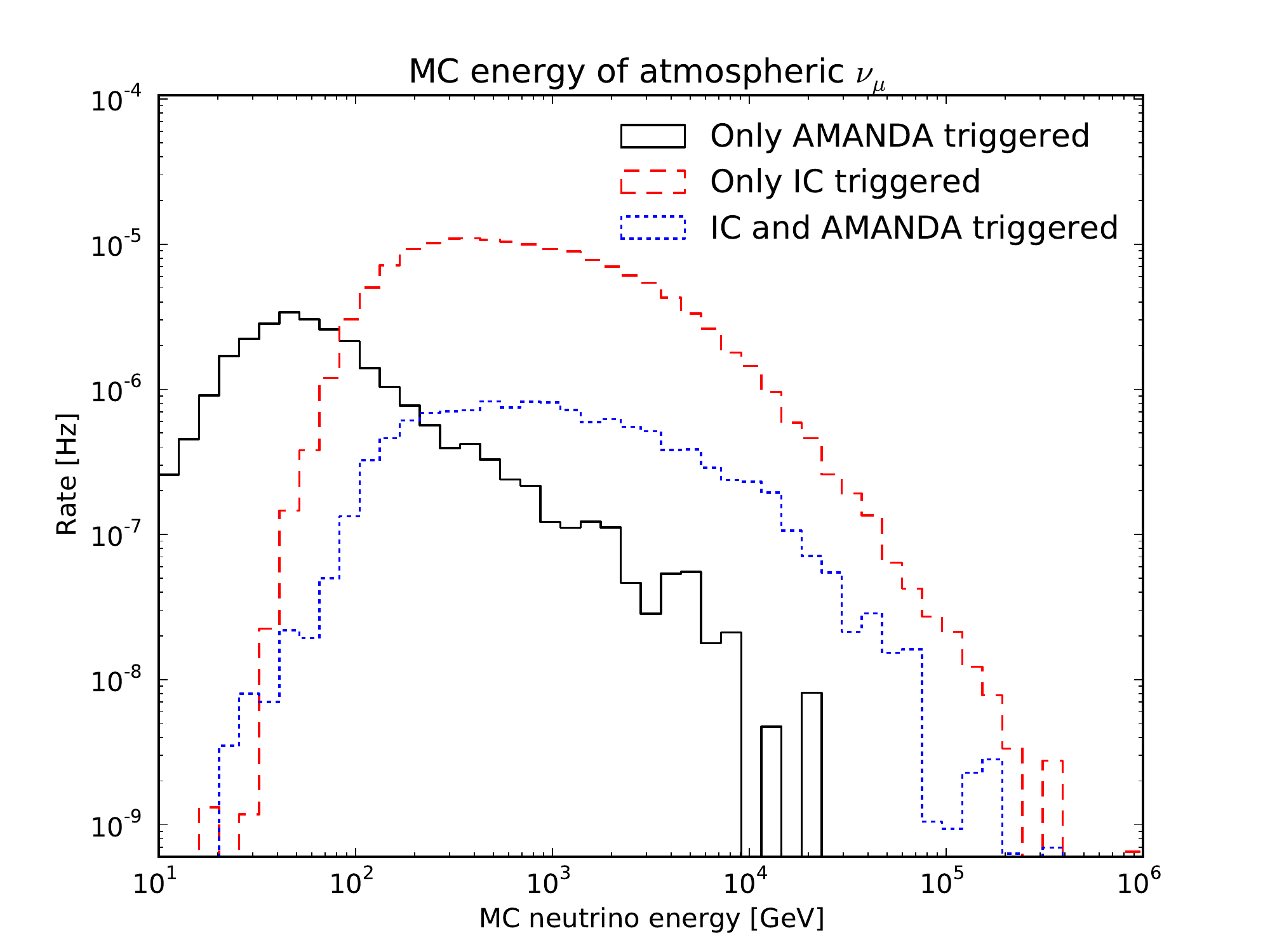}
\caption{Monte Carlo neutrino energy distribution at final level of Analysis B, of the events triggering only AMANDA--II (full line), only IceCube (dashed line) 
or both detectors (dotted line).}
\label{fig:triggers}
\end{figure}

\subsection{\label{subsec:analysis_b} ANALYSIS B\\ (AMANDA--IceCube combined analysis)}
   The data set used in this analysis comprises a total of 1.7$\times$10$^{10}$ events collected between the 
middle of April and end of September 2008, plus the period  March 20 -- April 1 2009. Only runs where both AMANDA--II and IceCube 
were active were used, which corresponds to a total livetime of 149 days when the Sun was below the horizon. The surface geometry of AMANDA--II/IceCube--40 
used in the analysis is shown in Figure~\ref{fig:geometry}. The denser AMANDA--II array plays the role of a low energy detector, as 
can be seen in Figure~\ref{fig:triggers} which shows the energy distribution of neutrinos triggering the AMANDA--II array and the 
IceCube array, at final cut level. We have therefore simplified  the analysis with respect to the approach taken in analysis A, and optimized the cuts 
for two energy regions, low neutrino energies and high neutrino energies. 
We used 100~GeV neutralinos annihilating into $b\bar{b}$ as a benchmark for the low--energy optimization, and 1000~GeV neutralinos annihilating 
into $W^+W^-$ for the high--energy optimization. The optimization that was finally used on each WIMP model was decided at the end of the analysis,  
based on which one achieved the best sensitivity for the given model.

\begin{figure*}[t]
\begin{minipage}[t]{\linewidth}
\includegraphics[width=0.45\linewidth]{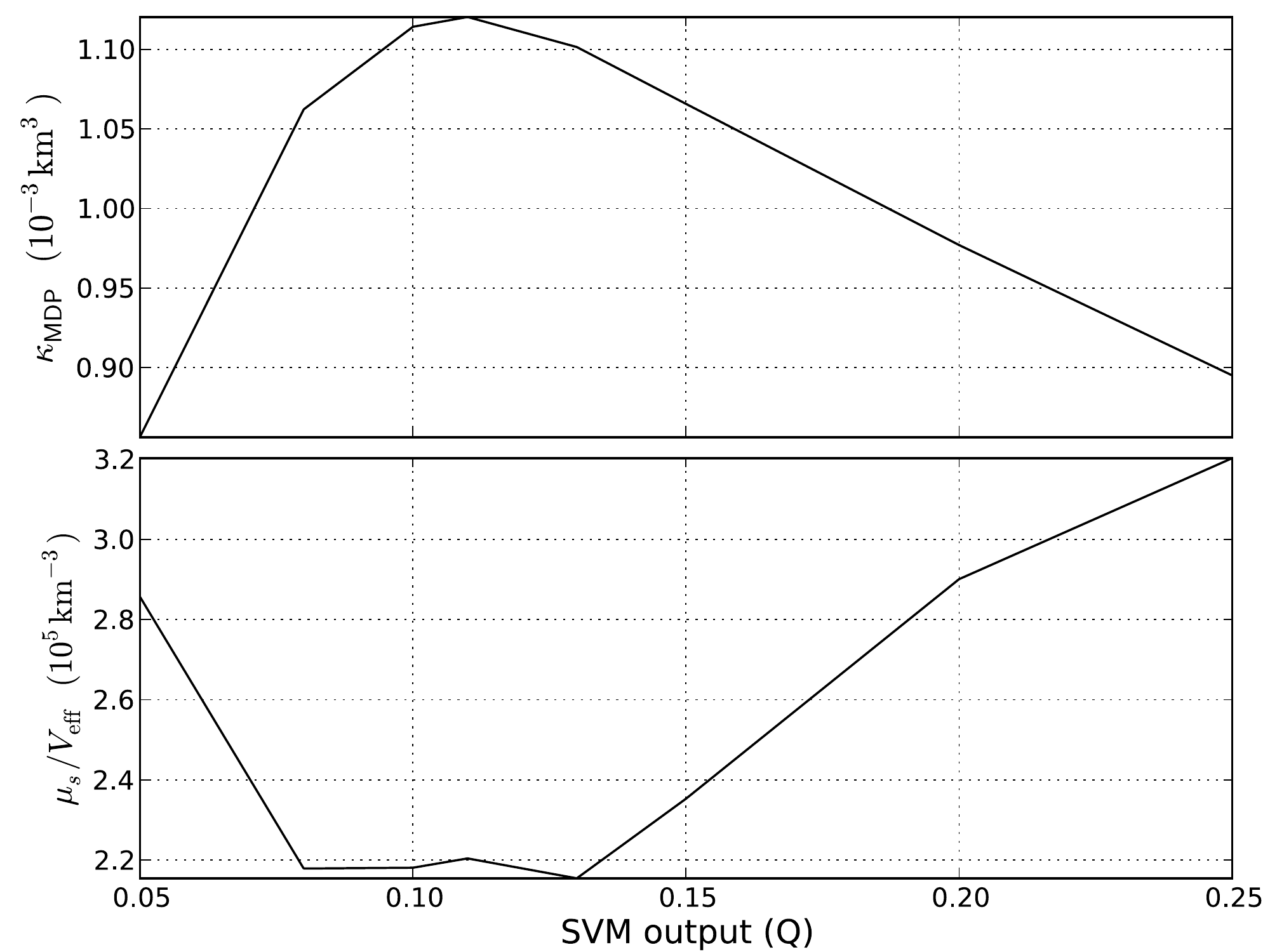}
\hspace*{0.5cm}
\includegraphics[width=0.45\linewidth]{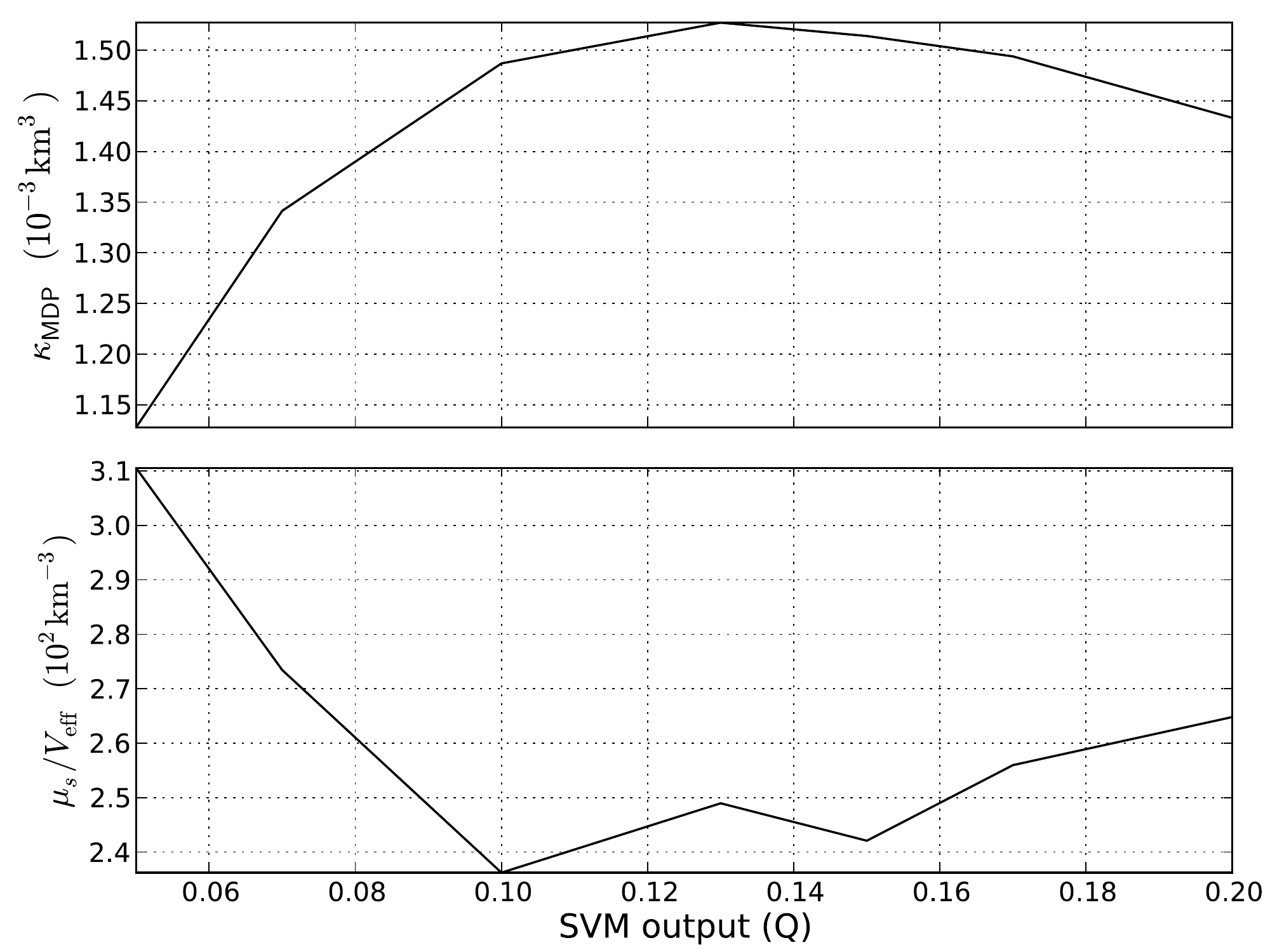}
\caption{Figure of merit of the model discovery potential (upper panels) and sensitivity (lower panels) as a function of 
SVM output of Analysis B. The left plots correspond to the low--energy analysis optimization  and the 
right plots to the high--energy optimization.}
\label{fig:mdp-sens}
\end{minipage}
\end{figure*}

\subsubsection{\label{subsec:ana_B} Data analysis}
 Data analysis proceeded on slightly different lines for the low--energy and high--energy streams. Events that had hits in the AMANDA--II 
detector were reconstructed with the JAMS first--guess reconstruction and then with an iterative likelihood fit. For events with hits only 
in the IceCube strings, a simple line--fit~\cite{Ahrens:04a} proved to work well and it was used as the first--guess.  A series of straight cuts on event quality 
and track direction were performed in order to reduce the amount of data while keeping as much of the signal as possible, before using a multivariate 
classifier for the final separation. The cut variables used at this stage were slightly different between the low--energy and the high--energy streams, 
reflecting the different geometry and size of the AMANDA--II and IceCube detectors, as well as the fact that we want to use IceCube as a veto for the 
low--energy stream. Events classified as belonging to the low--energy stream were required to have an AMANDA--II trigger, a reconstructed zenith angle larger 
than 90$^\circ$, at least 25\% of direct hits (with a minimum of four direct hits), a distance between the first and last direct hit of at least 25~m, 
and less than five hits in any of the IceCube strings. Events in the high--energy stream were also required to have a reconstructed zenith angle larger 
than 90$^\circ$ and smaller than 120$^\circ$, a reduced log--likelihood smaller than 6.5, a difference between the angles of the likelihood reconstruction and the 
first--guess fit smaller than 40$^\circ$, at least 3 direct hits, at least 2 strings with direct hits and a distance between the first and last direct 
hit of at least 141~m.  These cuts reduced the data and atmospheric muon simulation by a factor of about 2000 in each case, while just 
reducing the low--energy and high--energy signal streams by a factor of 9 and 2.4, respectively. \par
 After these straight cuts, a support vector machine (SVM)~\cite{SVM} was used for the final classification of the events. 
%Support Vector Machines are event classifiers which try to linearize the separation between two classes of events by mapping the input parameter space 
%into a higher dimensional space where a hyperplane is sought that maximizes the separation of the two classes. The easily separable events will be 
%classified and further ignored, while classification proceeds focusing only on the subset close to the hyperplane, the \emph{support vector}. 
% As any other classifier, SVMs need to be trained on a given signal and background sample. 
We trained two SVMs independently, one for the low--energy sample and another one for the high--energy sample. The signal training samples consisted of 50\% 
of the simulated signal samples considered in each stream. Data when the Sun is above the horizon were used as the background sample. The choice of variables 
to use in the SVM was done iteratively: from an original group of 35 variables, the SVM was trained on all variables but one at a time. A variable for which 
the SVM performance did not worsen when removed, was discarded. A set of 12 variables were identified by this method as useful for signal and background discrimination 
in the low--energy stream, and ten for the high--energy stream. As in Analysis A, the variables are related to the event topology and the quality of 
the track reconstruction. Three of the variables were common to both streams: the zenith difference between the reconstructed track and the 
position of the Sun, the angular difference between the reconstructed track and a line joining the centers of gravity of the first and the last (in time) 
quartiles of the hits, and the horizontal distance from the detector center to the center of gravity of the hits. Among the other variables 
used in a particular stream are the number of hit strings, the angular resolution of the track fit and the smoothness of the distribution of direct 
hits projected along the track direction. Details of the data analysis and the complete list of the variables used can be found in Ref.~\cite{Olle:11a}. 
The SVM classifies the given data as background or signal according to a continuous parameter, $Q$, that takes values between 0 (signal--like) and 
1 (background--like). %Figure~\ref{fig:svm} shows the distribution of the SVM output corresponding to the two analysis streams of Analysis B. 
A cut value of $Q=0.1$ was used for both data sets, and it is justified in Section~\ref{sec:results} below.  A fraction of 2.2$\times$10$^{-3}$ 
of the data in the low--energy analysis and 4.1$\times$10$^{-3}$ of the data in the high--energy analysis remain after this cut, 
while 24\% and 39\% of the signal are retained, respectively. No further cuts were applied after the SVM classification. 

\subsubsection{\label{subsec:sys_B}Systematic uncertainties}
 The uncertainties in analysis B were evaluated as in the previous analysis, by varying  the optical module sensitivity and the ice model in the Monte Carlo 
and evaluating the effect on the effective volume after all analysis steps. 
 The values for the uncertainties pertaining to theoretical inputs (neutrino--nucleon cross section, oscillation parameters and muon propagation in ice) were taken 
as for analysis A.\par
 With the deployment of IceCube, the ice properties at deeper depths than the AMANDA location needed to be evaluated, and a new ice model was developed that 
covered the whole IceCube volume and a better description of the layered structure of the ice. Two approaches were taken in updating the AMANDA 
ice properties modeling: extrapolating the AMANDA model based on new ice core data and using IceCube flasher data. The difference in the effective volume 
induced by these two approaches was taken as an estimation of the uncertainty in the analysis due to the ice optical properties. 
This uncertainty ranges from 2\% to 4\%, depending on the analysis stream, while the DOM sensitivity uncertainty lies between 12\% and 24\%. 
The uncertainties were added in quadrature to obtain the total uncertainty of each of the two streams of the analysis.

\begin{table*}[t]
\begin{ruledtabular}
\begin{tabular}{ccccccccc}
$m_{\widetilde{\chi}^{0}_{ 1}}$(GeV) & channel & ${\bar{\Psi}}$(deg) & $V_{\rm eff}$ (m$^3$) &
 $\mu_{\rm s}^{\rm 90}$ & $\Gamma_{\rm A}$(s$^{-1}$) & $\Phi_{\mu}$ (km$^{-2}$ y$^{-1}$)   &  $\sigma_{\chi p}^{\rm SD}$  (cm$^2$) & $\sigma_{\chi p}^{\rm SI}$ (cm$^2$) \\ 
\hline
50 &  $\tau^+\tau^-$ &  8.0 &  7.40$\times$10$^{4}$ &  10.8 &  8.11$\times$10$^{23}$ &  1.95$\times$10$^{4}$ &  1.86$\times$10$^{-39}$ &  7.55$\times$10$^{-42}$  \\
 &  $b\bar{b}$ &  13.1 &  5.49$\times$10$^{3}$ &  19.0 &  1.73$\times$10$^{26}$ &  1.81$\times$10$^{5}$ &  3.97$\times$10$^{-37}$ &  1.61$\times$10$^{-39}$  \\
\hline
100 &  $W^+W^-$ &  5.3 &  4.33$\times$10$^{5}$ &  7.8 &  1.19$\times$10$^{23}$ &  4.27$\times$10$^{3}$ &  9.60$\times$10$^{-40}$ &  2.41$\times$10$^{-42}$  \\
 &  $b\bar{b}$ &  9.0 &  4.81$\times$10$^{4}$ &  11.7 &  7.06$\times$10$^{24}$ &  2.30$\times$10$^{4}$ &  5.70$\times$10$^{-38}$ &  1.36$\times$10$^{-40}$  \\
\hline
250 &  $W^+W^-$ &  2.8 &  9.33$\times$10$^{6}$ &  6.2 &  2.99$\times$10$^{21}$ &  4.38$\times$10$^{2}$ &  1.41$\times$10$^{-40}$ &  1.95$\times$10$^{-43}$  \\
 &  $b\bar{b}$ &  4.7 &  3.35$\times$10$^{5}$ &  7.4 &  3.24$\times$10$^{23}$ &  3.76$\times$10$^{3}$ &  1.53$\times$10$^{-38}$ &  2.10$\times$10$^{-41}$  \\
\hline
500 &  $W^+W^-$ &  2.4 &  2.26$\times$10$^{7}$ &  5.3 &  9.23$\times$10$^{20}$ &  2.40$\times$10$^{2}$ &  1.70$\times$10$^{-40}$ &  1.73$\times$10$^{-43}$  \\
 &  $b\bar{b}$ &  3.4 &  1.50$\times$10$^{6}$ &  6.6 &  4.98$\times$10$^{22}$ &  1.24$\times$10$^{3}$ &  9.15$\times$10$^{-39}$ &  9.05$\times$10$^{-42}$  \\
\hline
1000 &  $W^+W^-$ &  2.2 &  3.23$\times$10$^{7}$ &  5.1 &  6.78$\times$10$^{20}$ &  2.04$\times$10$^{2}$ &  4.95$\times$10$^{-40}$ &  4.22$\times$10$^{-43}$  \\
 &  $b\bar{b}$ &  2.7 &  3.88$\times$10$^{6}$ &  5.9 &  1.39$\times$10$^{22}$ &  6.05$\times$10$^{2}$ &  1.01$\times$10$^{-38}$ &  8.67$\times$10$^{-42}$  \\
\hline
3000 &  $W^+W^-$ &  2.2 &  3.12$\times$10$^{7}$ &  5.2 &  1.01$\times$10$^{21}$ &  2.16$\times$10$^{2}$ &  6.56$\times$10$^{-39}$ &  4.97$\times$10$^{-42}$  \\
 &  $b\bar{b}$ &  2.6 &  8.00$\times$10$^{6}$ &  5.6 &  5.18$\times$10$^{21}$ &  3.70$\times$10$^{2}$ &  3.37$\times$10$^{-38}$ &  2.56$\times$10$^{-41}$  \\
\hline
5000 &  $W^+W^-$ &  2.2 &  3.05$\times$10$^{7}$ &  5.2 &  1.19$\times$10$^{21}$ &  2.14$\times$10$^{2}$ &  2.17$\times$10$^{-38}$ &  1.60$\times$10$^{-41}$  \\
 &  $b\bar{b}$ &  2.4 &  9.24$\times$10$^{6}$ &  5.7 &  4.32$\times$10$^{21}$ &  3.48$\times$10$^{2}$ &  7.81$\times$10$^{-38}$ &  5.78$\times$10$^{-41}$  \\
\end{tabular}
\end{ruledtabular}
\caption{\label{tab:results_B} ANALYSIS B: For each neutralino mass and annihilation channel the table shows: The  median of the space--angle distribution, 
${\bar{\Psi}}$, the effective volume, V$_{\rm eff}$,  the 90\% CL upper limit on 
the expected signal, $\mu_{\rm s}^{\rm 90}$, and the corresponding 90\% CL limits on the annihilation rate at the center of the Sun, $\Gamma_{\rm A}$, on the muon 
flux at the Earth, $\Phi_{\mu}$, and on the spin--dependent and spin--independent neutralino--proton cross sections, $\sigma_{\chi p}^{\rm SD}$  and  $\sigma_{\chi p}^{\rm SI}$.
The limits include systematic uncertainties.}
\end{table*}

\begin{figure*}[t]
\begin{minipage}[t]{\linewidth}
\includegraphics[width=0.45\linewidth,height=0.35\linewidth]{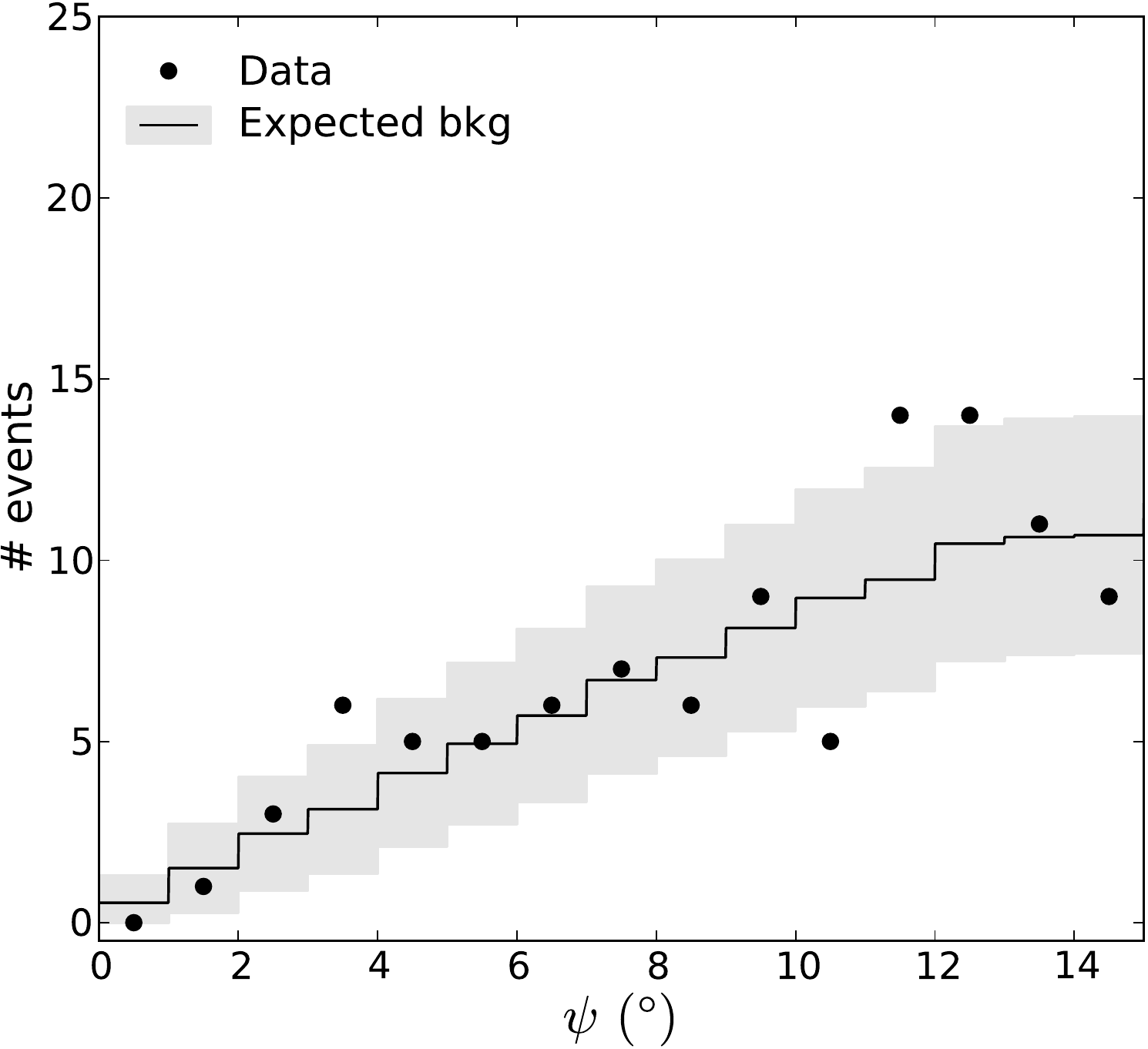}
\hfill
\includegraphics[width=0.45\linewidth,height=0.35\linewidth]{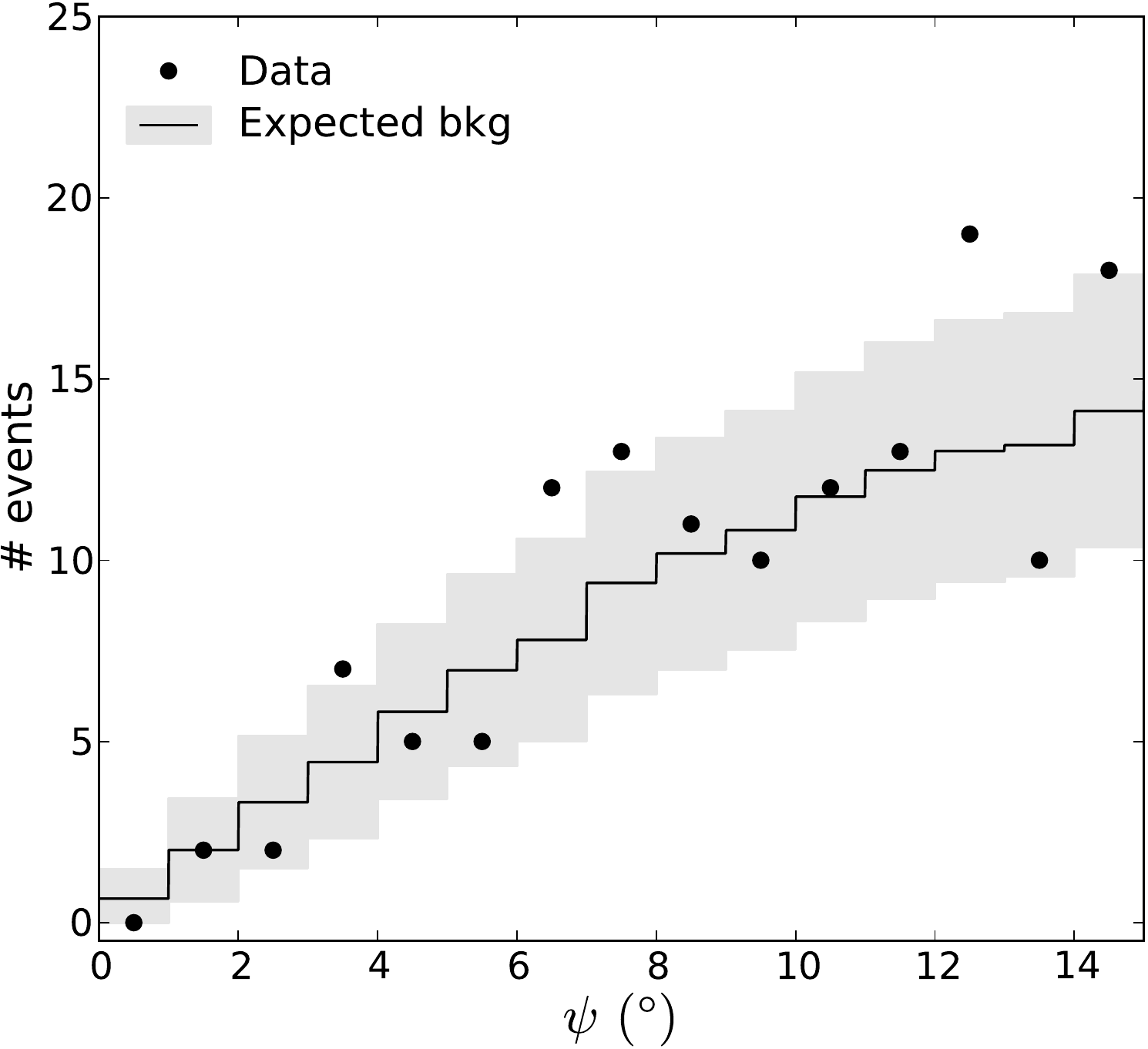}
\includegraphics[width=0.45\linewidth,height=0.35\linewidth]{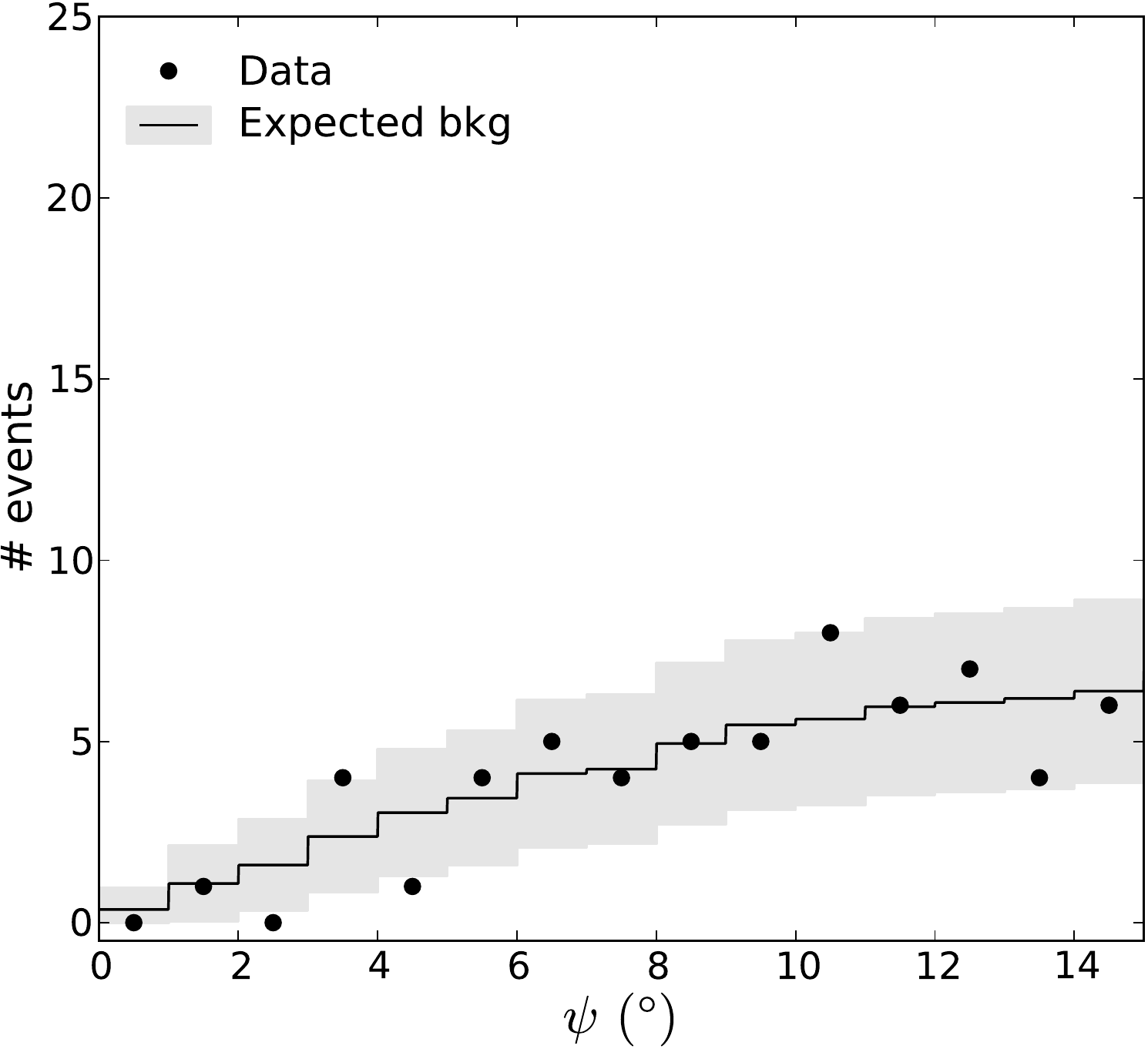}
\hfill
\includegraphics[width=0.45\linewidth,height=0.35\linewidth]{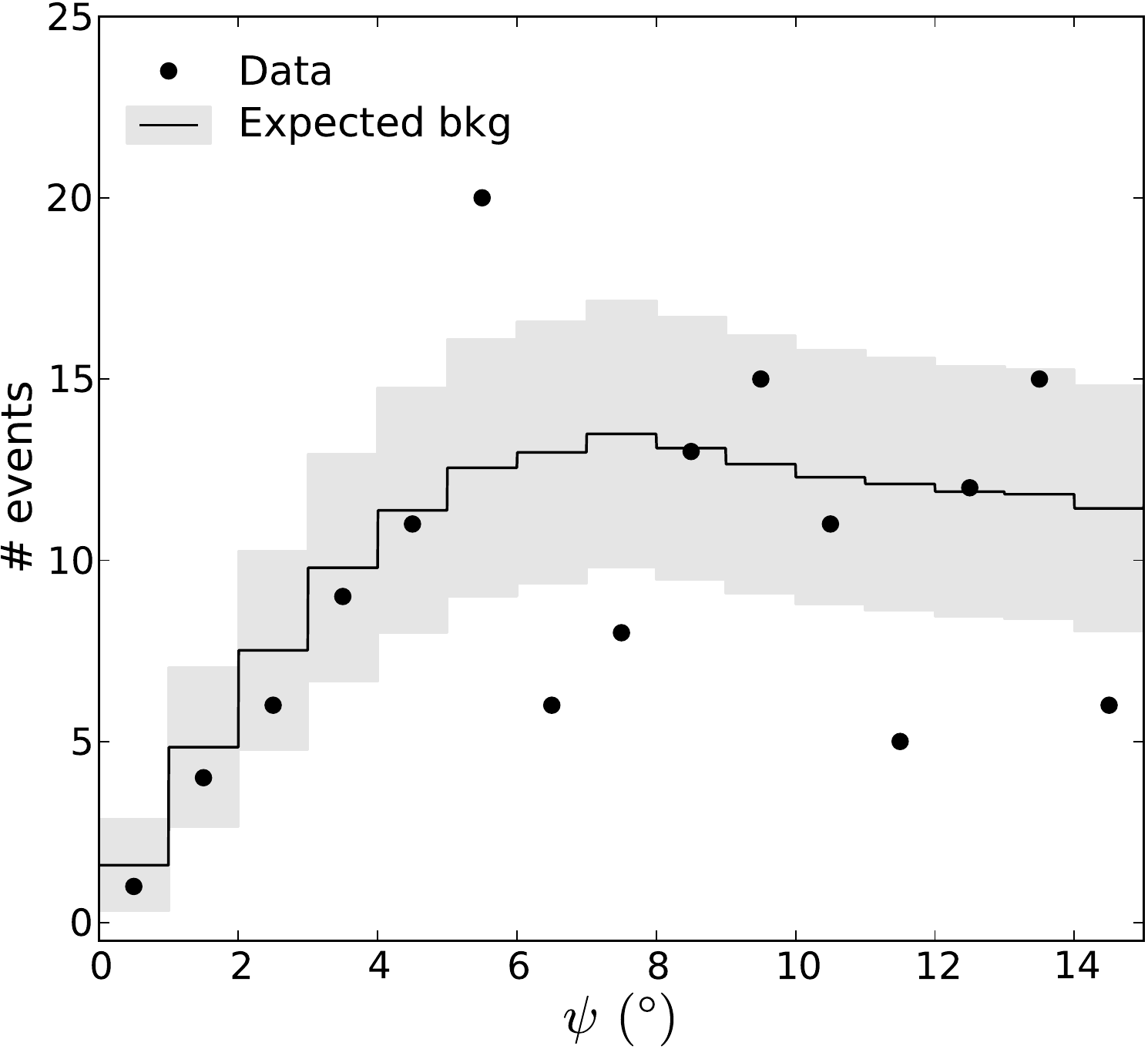}
\caption{The space angular distribution  with respect of the position of the Sun (first 15 degrees) of the 
remaining data events in two analyses, after all cuts (dots). The histogram in all plots represents the expected background distribution 
obtained from time--scrambled data in the same declination of the Sun, with the 1$\sigma$ Poisson uncertainty shown as the shaded area.
{\bf Upper row:} Analysis A: The left plot shows the case of optimizing the analysis for 100 GeV neutralinos and $b\bar{b}$ annihilation 
channel as an example of a low--energy optimization. The right plot shows the case of 1000 GeV neutralinos and $W^+W^-$ annihilation channel 
as an example of a high--energy optimization. 
{\bf Lower row:} Analysis B: The left plot shows the case of optimizing the analysis for 100 GeV neutralinos and $b\bar{b}$ annihilation 
channel (the low--energy optimization). The right plot shows the case of 1000 GeV neutralinos and $W^+W^-$ annihilation channel 
(the high--energy optimization)}
\label{fig:delta_phi}
\end{minipage}
\end{figure*}

\section{\label{sec:results}Results}
 The BDT classifiers of Analysis A were trained to maximize the model discovery potential (MDP) as defined in Ref.~\cite{Punzi:03a}, 
which measures the possible contribution of a signal when a given number of background events, $n_B$, are expected. Reference~\cite{Punzi:03a} gives 
an analytical parametrization  for the sensitivity of the experiment for a chosen discovery level at a given 
confidence level, 
\begin{equation}
\label{eq:punzi}
k_{\rm MDP} = \frac{\epsilon_S}{\frac{a^2}{8}+\frac{9b^2}{13}+a\sqrt{n_B}+\frac{b}{2}\sqrt{b^2+4a\sqrt{n_B}+4n_B}} 
\end{equation}
where $\epsilon_s$ represents the signal efficiency. This figure of merit was chosen to represent the strength of the signal 
flux needed for a 5$\sigma$ significance discovery ($a$=5) at 90\% confidence level ($b$=1.28). \par

The SVMs in Analysis B were trained to optimize the sensitivity (the median upper limit of the number of signal events at 90\% confidence level, 
$\bar{\mu}_{\mathrm{s}}^{90}$ divided by the detector effective volume, $V_{\mathrm{eff}}$) to the signal, although such optimization was found to be equivalent 
to maximizing the discovery potential. This is illustrated in Figure~\ref{fig:mdp-sens}, where lower panels  show the sensitivity, 
while the upper panels show the figure of merit of the discovery potential, both as a function of the cut on the SVM output $Q$. The left plot corresponds  
to the low--energy optimization and the right plot to the high--energy optimization. The figure shows that the chosen cut on the SVM ($Q=0.1$), 
based on the sensitivity plots, lies very close to the maximum of the discovery potential curves. \par

We use the shape of the space angle distribution of the final data samples with respect to the Sun to build a confidence interval for $\mu_{\rm s}$, 
the number of signal events compatible with the observed data distribution at a given confidence level, that we will choose to be 90\%. The 
plots in Figure~\ref{fig:delta_phi} show an example of the angular distribution of the remaining data after all cuts, compared with 
the angular distribution of the expected background obtained from time--scrambled data from the same declination of the Sun. Angles shown are space 
angles with respect to the Sun position, and the different plots correspond to Analysis A and Analysis B as indicated in the caption. \par

A probability density function for the space angle $\psi$ can be constructed as follows, 
\begin{equation}
 f(\psi|\mu_{\mathrm{s}}) = \frac{\mu_{\mathrm{s}}}{n_{\mathrm{obs}}}f_S(\psi)+ \left( 1 - \frac{\mu_{\mathrm{s}}}{n_{\mathrm{obs}}} \right) f_B(\psi)
\label{eq:compdf}
\end{equation}
representing the probability of observing a given angle $\psi$ when $\mu_{\mathrm{s}}$ signal events are present among the 
total number of observed events $n_{\mathrm{obs}}$. The functions $f_S(\psi)$ and $f_B(\psi)$ are the angular probability 
density functions of signal and background respectively, obtained by fitting and normalizing the corresponding angular distributions.
The likelihood of the presence of $\mu_{\mathrm{s}}$ signal events in an experiment that observed exactly $n_{\mathrm{obs}}$ events, can 
then be expressed as follows
\begin{equation}
\mathcal{L}(\mu_{\rm s}) = \prod_{i=1}^{n_{{\rm obs}}} f(\psi_i | \mu_{\mathrm s}),
 \label{eq:llh}
\end{equation}

 To define confidence intervals based on Eq.~(\ref{eq:compdf}) we use the likelihood--ratio test statistic 
$\mathcal{R}(\mu_{\rm s}) = \mathcal{L}(\mu_{\rm s})/\mathcal{L}(\widehat{\mu}_{\rm s})$ as proposed by Feldman and Cousins~\cite{Feldman:98a},
where $\widehat{\mu}_{\rm s}$ is the result of best fit to the observed ensemble of space angles. For each 
$\mu_{\rm s}$, $\mathcal{L}(\mu_{\rm s}) \leq \mathcal{L}(\widehat{\mu}_{\rm s})$ and $\mathcal{R}(\mu_{\rm s}) \leq 1$. Through a series of 
pseudo experiments with varying values of $\mu$, a critical value $\mathcal{R}^{90}(\mu_{\rm s})$ 
can be found such as ln$\mathcal{R}(\mu_{\rm s})\geq$ln$\mathcal{R}^{90}(\mu_{\rm s})$ for 90\% of the cases. A confidence interval 
[$\mu_{\rm low}$, $\mu_{\rm up}$] at 90\% confidence level can then be calculated as 
[$\mu_{\rm low}$, $\mu_{\rm up}$]=\{$\mu$ $\arrowvert$ ln$\mathcal{R}(\mu_{\rm s})\geq$ln$\mathcal{R}^{90}(\mu_{\rm s})$\}. \par

 In the absence of a signal, $\mu_{\rm up}$ is the 90\% confidence level limit on the number of signal 
events, $\mu_{\rm s}^{\rm 90}$. This limit can be directly transformed into a limit on the volumetric rate of 
neutrino interactions in the detector due to a signal flux, $\Gamma_{\nu \rightarrow \mu}$,  since we would 
expect a number $\mu_{\rm s} = \Gamma_{\nu \rightarrow \mu}  V_{\rm eff} \cdot t_{\rm live} $ of neutrinos in the 
livetime $t_{\rm live}$. Therefore 
\begin{equation}
\Gamma_{\nu \rightarrow \mu} \leq \Gamma_{\nu \rightarrow \mu}^{90} = \frac{\mu_{\rm s}^{\rm 90}}{V_{\rm eff}\cdot t_{{\rm live}}}
 \label{eq:convrate}
\end{equation}

This volumetric interaction rate is directly proportional to the neutralino annihilation rate in the Sun, $\Gamma_{\rm A}$, through 
\begin{multline}
%\begin{split}
\Gamma_{\nu \rightarrow \mu} = \frac{\Gamma_{\rm A}\, \rho_{\rm N}}{4\pi D_{\odot}^2}\int_0 ^{\infty} dE_{\nu} \, \sigma_{\nu N} 
\left( E_{\mu} \geq E_{\mu}^{\rm thr} \mid E_{\nu} \right)\\
 \sum_{i}P_{\rm osc}(\mu,i) \sum_{\rm K} B_{\rm K}  \left( \frac{dN_{i}^{\rm K}}{dE_{\nu}} \right)
%\end{split}
\label{eq:gammaA}
\end{multline}
where $D_{\odot}$ is the distance to the Sun, $\sigma_{\nu N}$ the neutrino--nucleon cross section (above a 
given muon energy threshold, $E_{\mu}^{\rm thr}$, taken as 10 GeV), $\rho_{\rm N}$ the nucleon number density of 
the detector medium, $P_{\rm osc}(\mu,i)$ the probability that a produced neutrino of flavour $i$ oscillates 
to flavour $\mu$ before reaching the detector (including the probability that the neutrino escapes the 
dense solar interior), $B_{\rm K}$ is the branching ratio for annihilation into channel K, and $dN_{i}^{\rm K}/dE_{\nu}$ 
the number of neutrinos of flavour $i$ produced per annihilation and unit of energy from channel K. 
The branching ratios to a given channel are the only unknown quantities in the above equation. They depend on several unknown SUSY parameters, i.e. the composition and the mass of the neutralino. 
To be able to make concrete predictions and simplify the way the results are presented, we have performed the analysis on 
two annihilation channels per simulated mass, $W^+W^-$ (or $\tau^+ \tau^-$ if $m_{\chi} < m_{\rm W}$) and $b\bar{b}$, 
assuming 100\% branching ratio to each channel in turn. For a given mass, these channels produce the hardest and softest 
neutrino energy spectra respectively and are taken as representative of the range of possible outcomes if nature chose the 
MSSM neutralino as dark matter.\par
\begin{table*}[t]
\begin{ruledtabular}
\begin{tabular}{ccccccc}
 $m_{\widetilde{\chi}^{0}_{1}}$(GeV) & channel &
 $\mu_{\rm s}^{\rm 90}$ & $\Gamma_{\rm A}$(s$^{-1}$) & $\Phi_{\mu}$ (km$^{-2}$ y$^{-1}$)   &  $\sigma_{\chi p}^{\rm SD}$  (cm$^2$) & $\sigma_{\chi p}^{\rm SI}$ (cm$^2$) \\
   &  & &  & & &  \\
\hline
50 &  $\tau^+\tau^-$ &  20.1 &  6.83$\times$10$^{23}$ &  1.64$\times$10$^{4}$ &  1.57$\times$10$^{-39}$ &  6.26$\times$10$^{-42}$  \\
 &  $b\bar{b}$ &  25.5 &  1.45$\times$10$^{32}$ &  1.13$\times$10$^{5}$ &  2.47$\times$10$^{-37}$ &  1.00$\times$10$^{-39}$  \\
\hline
100 &  $W^+W^-$ &  19.4 &  8.75$\times$10$^{22}$ &  3.15$\times$10$^{3}$ &  7.07$\times$10$^{-40}$ &  1.67$\times$10$^{-42}$  \\
 &  $b\bar{b}$ &  23.7 &  7.69$\times$10$^{29}$ &  1.83$\times$10$^{4}$ &  4.53$\times$10$^{-38}$ &  1.07$\times$10$^{-40}$  \\
\hline
250 &  $W^+W^-$ &  15.4 &  2.73$\times$10$^{21}$ &  3.99$\times$10$^{2}$ &  1.28$\times$10$^{-40}$ &  1.74$\times$10$^{-43}$  \\
 &  $b\bar{b}$ &  27.1 &  3.82$\times$10$^{23}$ &  4.42$\times$10$^{3}$ &  1.80$\times$10$^{-38}$ &  2.39$\times$10$^{-41}$  \\
\hline
500 &  $W^+W^-$ &  10.9 &  7.13$\times$10$^{20}$ &  1.86$\times$10$^{2}$ &  1.31$\times$10$^{-40}$ &  1.32$\times$10$^{-43}$  \\
 &  $b\bar{b}$ &  12.7 &  3.00$\times$10$^{22}$ &  7.50$\times$10$^{2}$ &  5.54$\times$10$^{-39}$ &  5.46$\times$10$^{-42}$  \\
\hline
1000 &  $W^+W^-$ &  6.8 &  3.43$\times$10$^{20}$ &  1.03$\times$10$^{2}$ &  2.50$\times$10$^{-40}$ &  2.12$\times$10$^{-43}$  \\
 &  $b\bar{b}$ &  14.0 &  1.08$\times$10$^{22}$ &  4.69$\times$10$^{2}$ &  7.86$\times$10$^{-39}$ &  6.56$\times$10$^{-42}$  \\
\hline
3000 &  $W^+W^-$ &  9.3 &  6.34$\times$10$^{20}$ &  1.36$\times$10$^{2}$ &  4.13$\times$10$^{-39}$ &  3.10$\times$10$^{-42}$  \\
 &  $b\bar{b}$ &  9.9 &  3.05$\times$10$^{21}$ &  2.17$\times$10$^{2}$ &  1.98$\times$10$^{-38}$ &  1.49$\times$10$^{-41}$  \\
\hline
5000 &  $W^+W^-$ &  8.7 &  7.11$\times$10$^{20}$ &  1.27$\times$10$^{2}$ &  1.28$\times$10$^{-38}$ &  9.36$\times$10$^{-42}$  \\
 &  $b\bar{b}$ &  9.4 &  2.34$\times$10$^{21}$ &  1.89$\times$10$^{2}$ &  4.25$\times$10$^{-38}$ &  3.09$\times$10$^{-41}$  \\
\end{tabular}
\end{ruledtabular}
\caption{\label{tab:results_C} Results from the combination of Analysis A, Analysis B and the results of Ref.~\cite{Abbasi:09b}, 
covering 1065 days of livetime between 2001 and early 2009. 
For each neutralino mass and annihilation channel the table shows: The  90\% CL upper limit on 
the expected signal, $\mu_{\rm s}^{\rm 90}$, and the corresponding 90\% CL limits on the annihilation rate at the center of the Sun, $\Gamma_{\rm A}$, on the muon 
flux at the Earth, $\Phi_{\mu}$, and on the spin--dependent and spin--independent neutralino--proton cross sections, $\sigma_{\chi p}^{\rm SD}$  and  $\sigma_{\chi p}^{\rm SI}$.
The limits include systematic uncertainties. }
\end{table*}

 Equivalently to Eq.~(\ref{eq:gammaA}), we can relate the annihilation rate and the muon flux at the 
detector, $\Phi_{\mu}$, above the energy threshold $E_{\mu}^{\rm thr}$ as 
\begin{multline} 
%\begin{split}
\Phi_{\mu}^{E_{\mu}  \geq E_{\rm thr}} = \frac{\Gamma_{\rm A}\, \rho_{\rm N}}{4\pi D_{\odot}^2} 
\int_{0}^{\infty} dE_{\nu} 
\int_{E_{\mu}^{\rm thr}}^{E_{\nu}} dE_{\mu}^{'}\,
%\int_{E_{\mu}'}^{E_{\mu}^{\rm thr}} dE_{\mu}^{''}\,\frac{dX}{dE_{\mu}^{''}}\,
X(E_{\mu}^{'})\\
\frac{d\sigma_{\nu}(E_{\nu},E_{\mu}^{'})}{dE_{\mu}^{'}} 
\sum_{i}P_{\rm osc}(\mu,i)\sum_{\rm K}B_{\rm K}\left( \frac{dN_{i}^{\rm K}}{dE_{\nu}} \right)
%\end{split}
\label{eq:muflux}
\end{multline}
where $d\sigma_{\nu}(E_{\nu},E_{\mu}^{'})/dE_{\mu}^{'}$ is the differential neutrino cross section 
for producing a muon with energy $E_{\mu}^{'}$ from a neutrino of energy $E_{\nu}$. The 
term $X(E_{\mu}^{'})$ denotes the range of the muon produced with energy $E_{\mu}^{'}$, and takes into 
account energy losses between the production point and the detector location. 
For each mass and annihilation channel considered, the integrals 
have been performed with {\texttt{WimpSim}}, providing a relationship between the experimentally obtained 
limit on $\Gamma_{\nu \rightarrow \mu}$ and derived limits on $\Gamma_{\rm A}$ and  $\Phi_{\mu}$.\par
 
 If we assume equilibrium between the capture and annihilation rates, the annihilation rate is 
proportional to the neutralino--proton scattering cross section, which drives the capture. Under the 
further assumption that the capture rate is fully dominated either by the spin--dependent (SD) or 
spin--independent (SI) scattering, we can extract conservative limits on either the SD or the SI 
neutralino--proton cross section from the limit on $\Gamma_{\rm A}$. We have followed the method described 
in Ref.~\cite{Wikstrom:09a} in order to extract limits on the SD and SI cross sections. This conversion 
introduces an additional systematic uncertainty in the calculation of the cross sections, due to the 
element composition of the Sun, effect of planets on the capture of halo WIMPs and nuclear form factors 
used in the capture calculations. These effects influence the SI and SD calculations differently, and 
are discussed in Ref.~\cite{Wikstrom:09a}.  The uncertainties introduced by the conversion are small 
for the spin--dependent cross section, 2\%, and larger for the spin--independent cross section, 25\%. 
We have added these contributions in quadrature in order to obtain the total systematic uncertainty 
on these quantities that we use in this work.  We just note that recent studies favour a somewhat 
higher value of the dark matter density, $\rho_0$, closer to 0.4~GeV/cm$^3$~\cite{Catena:10a}, than 
the standard value of 0.3 GeV/cm$^3$ assumed in our conversion. Since variations in the dark matter 
density translate inversely into the calculated WIMP--proton cross section, the conservative 
assumption on $\rho_0$ makes our limits conservative as well. A recent discussion on the effects of 
uncertainties on the structure of the dark matter halo, the dark matter velocity dispersion 
and the effect of the gravitational influence of planets on the capture rate of local dark matter 
has been presented in detail in Ref.~\cite{Rott:11a}. \par

 The results on the quantities described above for each neutralino mass and annihilation channel 
considered are presented in Table~\ref{tab:results_A} for Analysis A and in Table~\ref{tab:results_B} 
for Analysis B. The tables show  the median of the space angle distribution with respect to the Sun, 
the effective volume for each analysis optimization, the 90\% CL limits on the expected signal, on the 
annihilation rate at the center of the Sun, on the muon flux at the Earth, and on the spin--dependent 
and spin--independent neutralino--proton cross sections.

\section{\label{sec:combination} Combination of results}
 Given that the data samples used in the analyses presented in this paper are independent, we can 
combine the results in a statistically sound way. Since IceCube has already published an analysis using 
the same method on another independent data set, the data collected with the 22--string detector in 2007, 
we can combine these three analyses and cover a total livetime of 1065 days, stretching from March 
2001 to April 2009. We use the combined likelihood constructed from the likelihoods of the three analyses,
\begin{multline}
%\begin{split}
\mathcal{L}(\mu_{\rm s}) =  \mathcal{L}_1(\mu_{\rm s}\omega_1) \mathcal{L}_2(\mu_{\rm s}\omega_2) \mathcal{L}_3(\mu_{\rm s}\omega_3) = \\
\prod_{i=1}^{n^{\rm obs}_{1}} f_1(\psi_{i,1} | \mu_{\rm s}\omega_1) \times \prod_{i=1}^{n^{\rm obs}_{2}} f_2(\psi_{i,2} | \mu_{\rm s}\omega_2) \times 
\prod_{i=1}^{n^{\rm obs}_{3}} f_3(\psi_{i,3} | \mu_{\rm s}\omega_3)
%\end{split}
 \label{eq:llh_combined}
\end{multline}
where  $\mu_{\rm s}$ is now weighted by the livetime $t_{{\rm live},i}$ and effective volume $V_{{\rm eff},i}$ of each analysis through the weights $\omega_i$, defined as  
\begin{equation}
\omega_i\,=\,\frac{V_{{\rm eff},i}\,t_{{\rm live},i}}{\sum_{j=1}^{3} V_{{\rm eff},j}\,t_{{\rm live}, j}}
 \label{eq:mu_combined}
\end{equation}
 We have chosen to show conservative limits and therefore we have used in Eq.~\ref{eq:mu_combined} the effective volumes of each analysis reduced by 
its 1$\sigma$ systematic uncertainty. 
 A 90\% confidence limit for $\mu_{\rm s}$ is then obtained using the same procedure as for the single--analysis case explained in Section~\ref{sec:results}. 
The combined limit on $\Gamma_{\nu \rightarrow \mu}$ is now given by
\begin{equation}
\Gamma_{\nu \rightarrow \mu} = \frac{\mu_{\rm s}^{\rm 90}}{\sum_{j=1}^{3} V_{{\rm eff},j}\,t_{{\rm live},j}}
 \label{eq:convrate_combined}
\end{equation}
 
\begin{figure}[t!]
\includegraphics[height=0.68\linewidth,width=\linewidth]{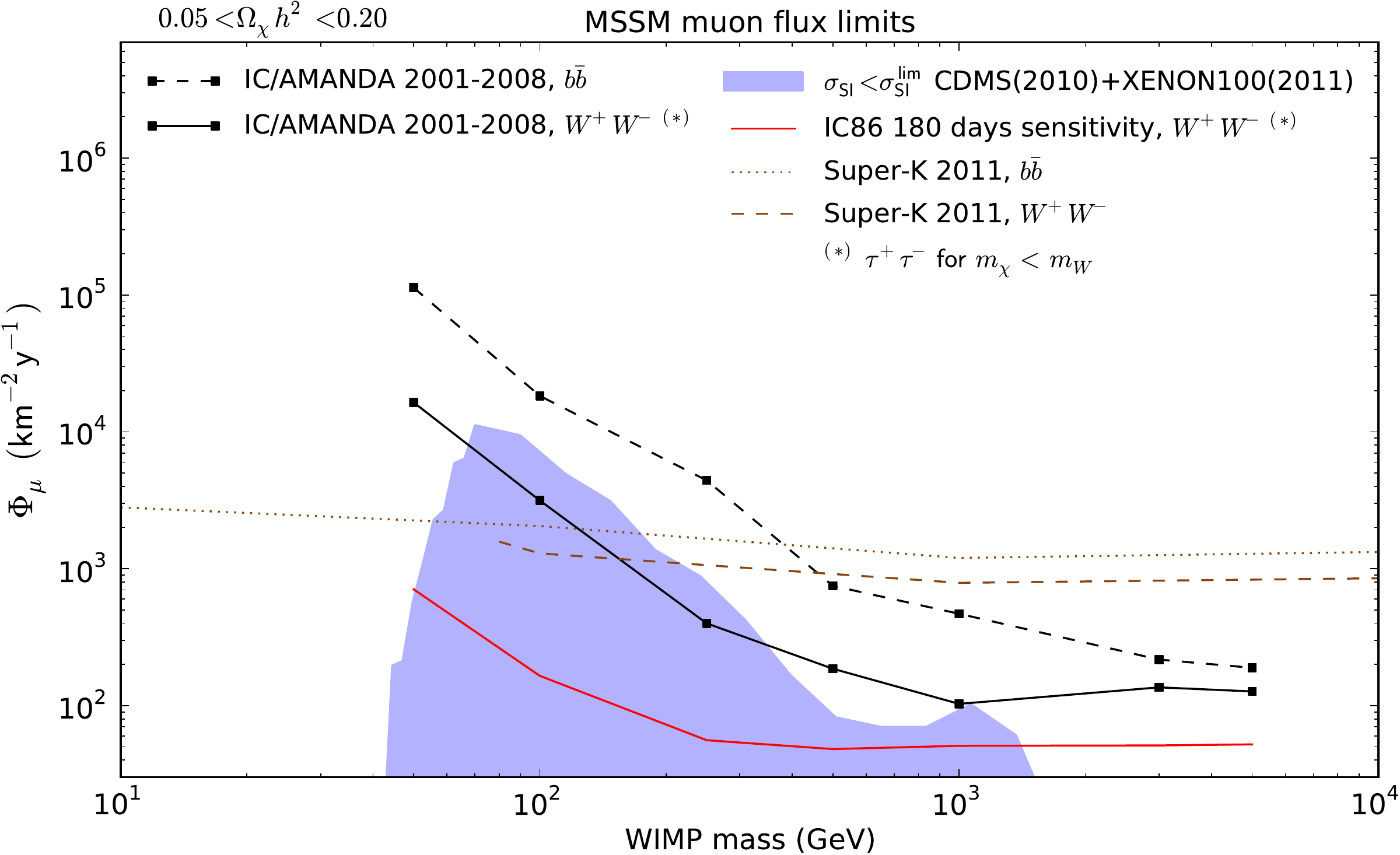}
\caption{ 90\% confidence level upper limits on the muon flux from the Sun, $\Phi_{\mu}$, from neutralino annihilations as a function of neutralino mass. 
The results from the analyses presented in this paper are shown as the black dots joined with lines to guide the eye (solid and dashed for the $W^+W^-$ 
and the $b\bar{b}$ annihilation channels respectively). The shaded area 
represents the allowed MSSM parameter space taking into account current accelerator, cosmological and direct dark matter search constrains. 
The red curve shows the expected sensitivity of the completed IceCube detector. Super--K results~\cite{SuperK:11a} are also shown for comparison.}
\label{fig:mu_limits}
\end{figure}

\begin{figure*}[t]
\begin{minipage}[t]{\linewidth}
\includegraphics[height=0.33\linewidth,width=0.48\linewidth]{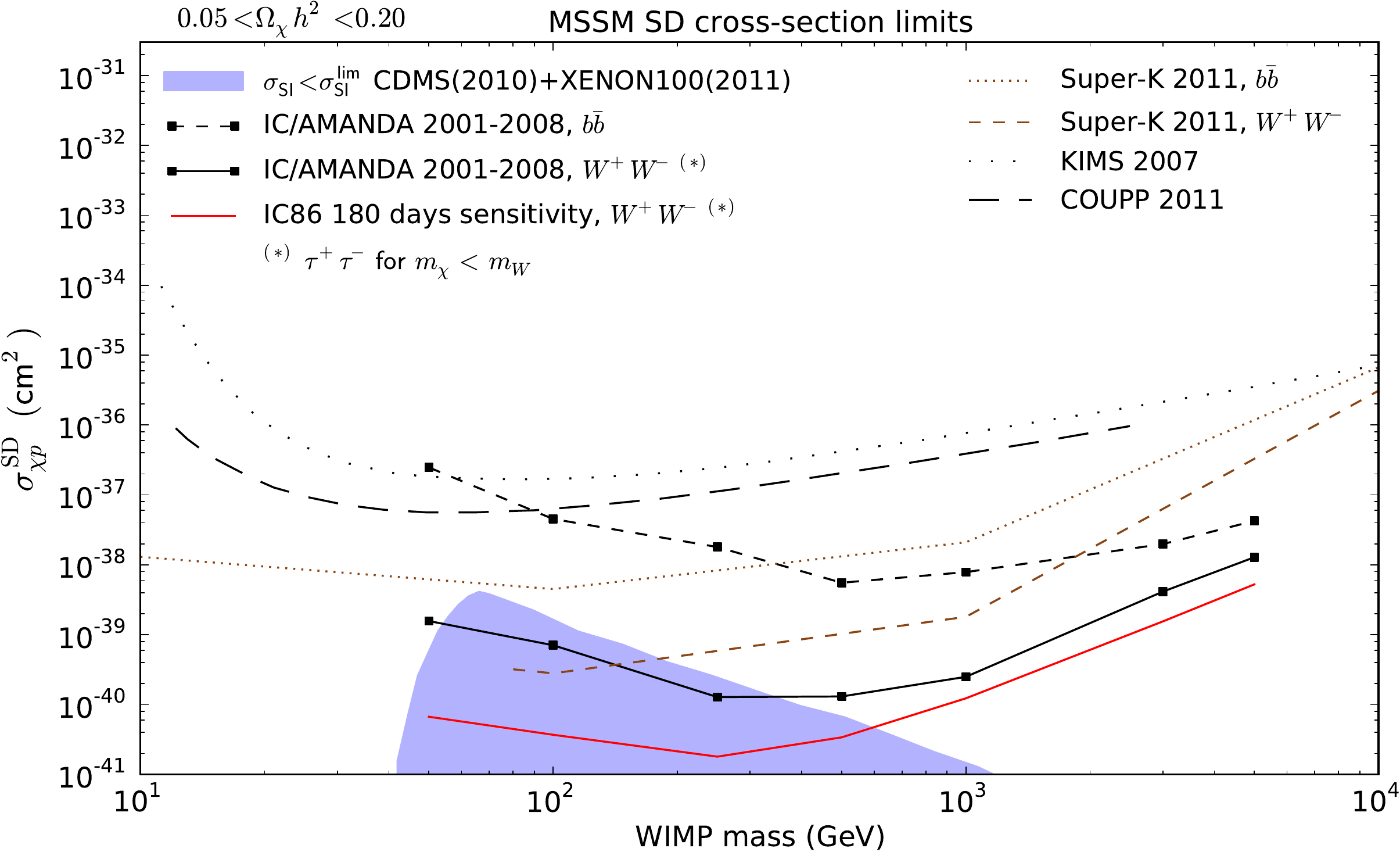}
\hfill
\includegraphics[height=0.33\linewidth,width=0.48\linewidth]{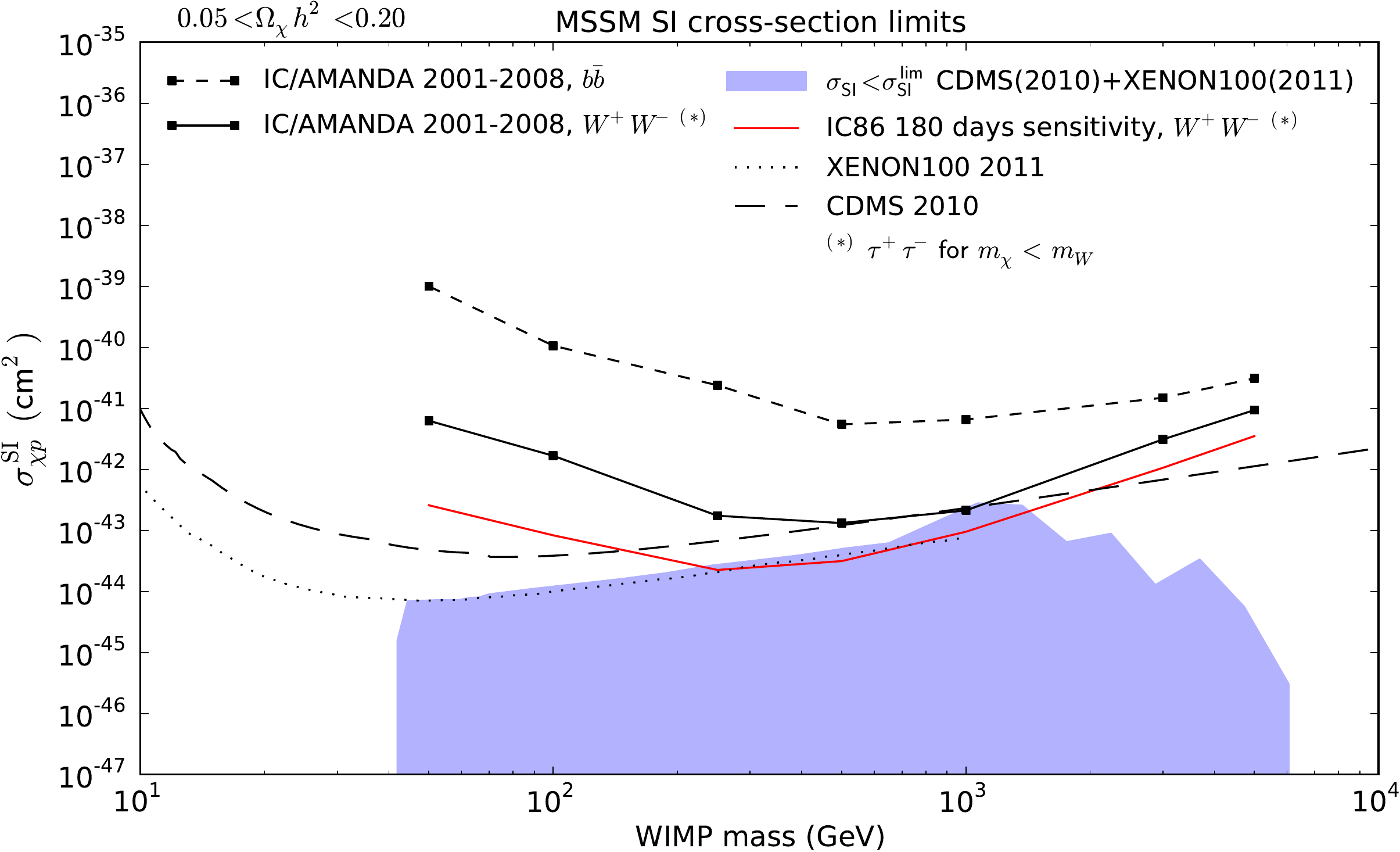}
\caption{{\bf Left:} 90\% confidence level upper limits on the spin--dependent neutralino--proton cross section, $\sigma_{\chi p}^{\rm SD}$, as a function 
of neutralino mass. The results from the analyses presented in this paper are shown as the black dots joined with lines to guide the eye (solid and 
dashed for the $W^+W^-$ and the $b\bar{b}$ annihilation channels respectively). The shaded area represents the allowed MSSM parameter space taking 
into account current accelerator, cosmological and direct dark matter search constrains. The red curve shows the expected sensitivity of the completed 
IceCube detector.  Results from Super--K~\cite{SuperK:11a}, KIMS~\cite{KIMS:07a} and COUPP~\cite{COUPP:11a} are also shown for comparison. 
{\bf Right:} 90\% confidence level upper limits on the spin--independent neutralino--proton cross section, 
$\sigma_{\chi p}^{\rm SI}$, as a 
function of neutralino mass. The results from the analyses presented in this paper are shown as the black dots joined with lines to guide the eye 
(solid and dashed for the $W^+W^-$ and the $b\bar{b}$ annihilation channels respectively). The shaded area represents the allowed MSSM parameter space 
taking into account current accelerator, cosmological and direct dark matter search constrains. The red curve shows the expected sensitivity of the 
completed IceCube detector. Results from  CDMS~\cite{CDMS:11a} and Xenon~\cite{Xenon:11a}  are also shown for comparison.
}
\label{fig:Xsec_limits}
\end{minipage}
\end{figure*}

The calculation of the combined limits on the annihilation rate and muon flux follows from $\Gamma_{\nu \rightarrow \mu}$ as in 
Eqs.~(\ref{eq:gammaA}) and~(\ref{eq:muflux}). The results of this procedure are shown in Table~\ref{tab:results_C} and in 
Figures~\ref{fig:mu_limits} and~\ref{fig:Xsec_limits}. The figures show the 90\% CL limits on the muon flux, on the 
spin--dependent and on the spin--independent neutralino--proton cross sections, compared with current limits from other experiments. 
The shaded area shows the allowed parameter space of the 7--parameter MSSM, obtained by a grid scan using {\texttt{DarkSusy}}~\cite{DarkSusy}. In order to choose only 
allowed models we have taken into account current experimental limits on the neutralino mass from LEP~\cite{PDG:10} and limits on the WIMP cross 
section from the CDMS~\cite{CDMS:11a} and XENON~\cite{Xenon:11a} direct detection experiments. We have allowed for a generous range of values of the dark matter 
relic density $\Omega_{\chi}h^2$ around the favored value of WMAP~\cite{WMAP:09a},  accepting models in the scan which predicted values of $\Omega_{\chi}h^2$ between 0.05 and 0.2.\par 

 An independent analysis using the point--source search techniques described in Ref.~\cite{Abbasi:09a} has been performed, using the Sun as another point source, 
and has been presented in Ref.~\cite{Braun:09a}. The analysis used the cuts developed for the point--source search without any further optimization.   
The only difference being that the estimated energy of the events was not included in the likelihood test in order to avoid the bias to 
the $E^{-2}$ spectrum used in the optimization of the point--source analysis.  Since this data set (2000--2006) practically overlaps with that of 
Analysis A, we have not included the results of such an approach in the combination presented above, but 
the results of the point--source analysis provide a useful confirmation of the robustness of the limits presented in this paper. \par

\section{\label{sec:outlook}Discussion}
 We have presented two independent analyses searching for neutralino dark matter accumulated in the center of the Sun 
using the AMANDA--II and the 40--string IceCube detectors, covering different data taking periods. We have combined the obtained 
90\% confidence level limits on the muon flux and neutralino--proton spin--dependent and spin--indepenent cross section with the previously published 
limits from the 22--string IceCube configuration, to obtain limits corresponding to a total livetime of 1065 days. We compare these results to the 
currently allowed MSSM--7 parameter space, obtained by a grid scan using {\texttt{DarkSusy}}. The limits on the muon flux are the most constraining so far for 
neutralino masses $m_{\chi^0}\gtrsim 100$ GeV. Assuming that the neutralinos constitute the bulk of dark matter in the Galaxy and that they annihilate 
producing a hard neutrino spectrum at the detector, the combined results of AMANDA--II and IceCube presented in this paper start to probe the allowed 7--parameter MSSM space.  
This comparison with the MSSM is, by construction of these analyses, limited to a model--by--model rejection/acceptance, but is not able to say anything about the 
relative probability of certain regions of the parameter space with respect to others, neither identify a best--fit model 
given the results obtained. This is the reason we have chosen to show the allowed MSSM space as a uniform shaded area, since individual model density in 
the plot would not carry any statistical meaning. Analyses that use experimental results to assign probabilities to the SUSY parameter space are becoming 
common.  A first evaluation of the possibilities that IceCube presents in probing the constrained MSSM, including individual model information, was done 
in Ref.~\cite{Trotta:10a}. A more elaborate analyses in this direction, including event energy and direction information, as well as constraints from 
direct dark matter search experiments and accelerator results in a global fit to the SUSY parameter space 
is being developed by the collaboration with data obtained with the 79--string detector. \par

Given that the Sun is essentially a proton target and that the muon flux at the detector can be related 
to the capture rate of neutralinos in the Sun, the IceCube limits on the spin--dependent neutralino--proton 
cross section are currently well below the reach of direct search experiments, proving that 
neutrino telescopes are competitive in this aspect. For the spin--independent limits, however, direct 
dark matter search experiments can be competitive due to the choice of target. Indeed, the latest results 
from the XENON100 collaboration~\cite{Xenon:11a}, using 100 days of livetime, have already produced 
stronger spin independent limits than those we present in this paper, as shown in the right plot of 
figure~\ref{fig:Xsec_limits}. However there is some complementarity between direct and 
indirect searches for dark matter given the astrophysical assumptions inherent to the calculations. Both 
methods are sensitive to opposite extremes of the velocity distribution of dark matter particles in the 
Galaxy (low--velocity particles are captured more efficiently in the Sun, high--velocity particles leave 
clearer signals in direct detection experiments), as well as presenting different sensitivity to the 
structure of the dark matter halo (a local void  or clump can deplete or enhance the possibilities for 
direct detection).\par

 The data set used in Analysis B covered the time until the decommissioning of the AMANDA--II detector 
in 2009. The denser configuration of the AMANDA--II strings was of key importance on increasing the 
sensitivity to low neutralino masses, while the sparsely spaced IceCube strings alone would have yielded 
a worse result. In order to supplant the role of AMANDA--II as a low--energy array, the IceCube 
collaboration has deployed the DeepCore array~\cite{Abbasi:11a}  in the clear South Pole ice, in the 
middle of the IceCube layout. DeepCore lies about 500~m deeper than AMANDA--II and its placement in 
the center of IceCube means that three layers of IceCube strings can be used as a veto to reject 
down--going atmospheric muons. The deployment of DeepCore was finalized in December 2010 and it is 
currently taking data embedded in the IceCube data acquisition system. DeepCore is expected to lower 
the energy threshold of IceCube to the {$\mathcal O$}(10 GeV) region, and therefore be an important 
asset in future dark matter searches with IceCube.

\begin{acknowledgments}
{\small 
We acknowledge the support from the following agencies:
U.S. National Science Foundation-Office of Polar Programs,
U.S. National Science Foundation-Physics Division,
University of Wisconsin Alumni Research Foundation,
the Grid Laboratory Of Wisconsin (GLOW) grid infrastructure at the University of Wisconsin - Madison, the Open Science Grid (OSG) grid infrastructure;
U.S. Department of Energy, and National Energy Research Scientific Computing Center,
the Louisiana Optical Network Initiative (LONI) grid computing resources;
National Science and Engineering Research Council of Canada;
Swedish Research Council,
Swedish Polar Research Secretariat,
Swedish National Infrastructure for Computing (SNIC),
and Knut and Alice Wallenberg Foundation, Sweden;
German Ministry for Education and Research (BMBF),
Deutsche Forschungsgemeinschaft (DFG),
Research Department of Plasmas with Complex Interactions (Bochum), Germany;
Fund for Scientific Research (FNRS-FWO),
FWO Odysseus programme,
Flanders Institute to encourage scientific and technological research in industry (IWT),
Belgian Federal Science Policy Office (Belspo);
University of Oxford, United Kingdom;
Marsden Fund, New Zealand;
Japan Society for Promotion of Science (JSPS);
the Swiss National Science Foundation (SNSF), Switzerland;
J.~P.~Rodrigues acknowledges support by the Capes Foundation, Ministry of Education of Brazil.
}
\end{acknowledgments}

\bibliographystyle{natbib}

\end{document}